\documentclass[a4paper,fleqn,usenatbib]{mnras}

\usepackage{newtxtext,newtxmath}

\usepackage[T1]{fontenc}
\usepackage{ae,aecompl}

\usepackage{graphicx}	
\usepackage{amsmath}	
\graphicspath{{figure/}} 
\usepackage{lipsum}
\usepackage{arydshln}
\usepackage{rotating}
\usepackage{graphicx}
\usepackage{geometry}
\usepackage{lscape}
\usepackage{url}
\usepackage{multicol}
\usepackage{multirow}
\usepackage{mathtools}
\usepackage{enumerate}
\usepackage{blindtext}
\usepackage{bm}
\usepackage{amsbsy}
\usepackage{fixmath}
\usepackage[english]{babel}
\usepackage{float}
\usepackage{caption}
\usepackage{gensymb}
\usepackage[utf8]{inputenc}
\usepackage{fixltx2e}
\usepackage{xcolor}

\def\simgt{\lower.5ex\hbox{$\; \buildrel > \over \sim \;$}}

\def\aap{A\&A}
\def\apj{ApJ}
\def\mnras{MNRAS}
\def\apjs{ApJS}
\def\aj{AJ}
\def\Msun{M_{\odot}}

\def\Zsun{Z_{\odot}}
\def\K{\rm{K}}
\def\nHtc{n_{\rm{H},t_{\rm c}}}

\title[SEAGLE-II: Strong lensing in EAGLE galaxy formation scenarios]{SEAGLE--II: Constraints on feedback models in galaxy formation from massive early-type strong lens galaxies}

\author[Mukherjee et al.]{Sampath\,Mukherjee$^{1,2}$\thanks{\href{mailto:sampath@astro.rug.nl}{\nolinkurl{sampath@astro.rug.nl}}},
L\'{e}on\,V.\,E.\,Koopmans$^{1}$, R.\,Benton\,Metcalf$^{3,4}$, \newauthor Cresenzo\, Tortora$^{5,6,1}$,  Matthieu\,Schaller$^{7}$, Joop\,Schaye$^{7}$, Georgios\,Vernardos$^{1}$, \newauthor Fabio\,Bellagamba$^{3,4}$
\\
\\
$^{1}$Kapteyn Astronomical Institute, University of Groningen, PO Box 800, 9700AV Groningen, The Netherlands\\
$^{2}$STAR Institute, Quartier Agora - All\'{e}e du six Ao$\hat{u}$t, 19c B-4000 Li\`ege, Belgium \\
$^{3}$Dipartimento di Fisica e Astronomia, Universit\`a di Bologna, via Gobetti 93/2, I-40129 Bologna, Italy\\
$^{4}$INAF -- Osservatorio di Astrofisica e Scienza dello Spazio di Bologna, via Gobetti 93/3, I-40129 Bologna, Italy\\
$^{5}$INAF -- Osservatorio Astronomico di Capodimonte, Salita Moiariello 16, 80131 - Napoli,
Italy\\
$^{6}$INAF -- Osservatorio Astrofisico di Arcetri, Largo Enrico Fermi 5, 50125, Firenze, Italy\\
$^{7}$Leiden Observatory, Leiden University, PO Box 9513, 2300 RA Leiden, The Netherlands\\
}

\date{Accepted XXX. Received YYY; in original form ZZZ}

\pubyear{2018} 

\begin{document}
\label{firstpage}
\pagerange{\pageref{firstpage}--\pageref{lastpage}}
\maketitle

\begin{abstract}
We use nine different galaxy formation scenarios in ten cosmological simulation boxes from the EAGLE suite of $\Lambda$CDM hydrodynamical simulations to assess the impact of feedback mechanisms in galaxy formation and compare these to observed strong gravitational lenses. To compare observations with simulations, we create strong lenses with  $M_\ast > 10^{11}$~M$_\odot$ with the appropriate resolution and noise level, and model them with an elliptical power-law mass model to constrain their total mass density slope. We also obtain the mass-size relation of the simulated lens-galaxy sample. We find significant variation in the total mass density slope at the Einstein radius and in the projected stellar mass-size relation, mainly due to different implementations of stellar and AGN feedback. We find that for lens selected galaxies, models with either too weak or too strong stellar and/or AGN feedback fail to explain the distribution of observed mass-density slopes, with the counter-intuitive trend that increasing the feedback steepens the mass density slope around the Einstein radius ($\approx$ 3-10 kpc). Models in which stellar feedback becomes inefficient at high gas densities, or weaker AGN feedback with a higher duty cycle, produce strong lenses with total mass density slopes close to isothermal (i.e.\ $-d\log(\rho)/d\log(r)\approx 2.0$) and slope distributions statistically agreeing with observed strong lens galaxies in SLACS and BELLS. Agreement is only slightly worse with the more heterogeneous SL2S lens galaxy sample. Observations of strong-lens selected galaxies thus appear to favor models with relatively weak feedback in massive galaxies.
\end{abstract}


\section{Introduction}\label{intro}

Large-scale numerical simulations have established the Cold Dark Matter (CDM) paradigm as a viable framework for galaxy formation (e.g.\ \citealt{davis1985,frenk1988}). The CDM model predicts that galaxies form in dark matter halos having a Navarro-Frenk-White (NFW) density profile (\citealt{navarro1996,navarro1997}) and predict the abundance and distribution of substructures within these halos (e.g. \citealt{gao2004,springel2010}). The physics of galaxy formation
, however, complicates the description of the matter distribution on small (several kpc) scales. Moreover, the central regions of CDM halos can  also be strongly modified by baryonic matter and their associated physical processes. 
Baryons settle into the centers of density concentrations due to dissipation, thereby modifying the inner DM slopes  (e.g. \citealt{Duffy2010,sonnenfeld2012,grilo2012,remus2013,cappellari2013a,CT2009,CT2010,Tortora14,Pontzen2014}). Because a complete analytic theory of baryonic physics is lacking, hydrodynamic simulations that include many physical processes have emerged as the dominant tool to study the complex non-linear interactions taking place during galaxy formation (e.g. \citealt{schaye2010,vogel2014,s15,Dubois2016,Hopkins2016}). State-of-the art hydrodynamical simulations with improved stellar and AGN feedback, for example, can reproduce the cosmic star formation history of the Universe and the galaxy stellar mass function. 
Hydrodynamic simulations are currently working only above certain mass and spatial resolutions, however, and physical processes on smaller scales are implemented via analytic prescriptions known as `sub-grid physics'. The impact of varying sub-grid physics prescriptions on large representative populations of stellar systems was first systematically explored in the `OverWhelmingly Large Simulations' project (OWLS; \citealt{schaye2010}), a suite of over fifty large cosmological hydrodynamical simulations with varying sub-grid physics. Calibration of sub-grid prescriptions to reproduce a limited number of observables has been explored extensively (\citealt{vogel2014,s15,c15,McCarthy2017}), showing that their exact parameterizations are very important. \\ 

Strong gravitational lensing is one of the most robust and powerful techniques to measure the total mass and its distribution in galaxies on kpc scales (\citealt{kochanek1991,koopmans2006}), allowing their inner structure and evolution over cosmic time to be studied in detail (\citealt{treu2006,treu2009,koopmans2006,koopmans2009,Dutton_Treu14}), independently of the nature of the matter or its dynamical state.  In particular, the mass density profile of massive lensing galaxies at $z>0.1$ can trace their formation and evolution mechanisms (e.g. \citealt{barnabe2009,barnabe2011}). The last two decades have seen major progress in observational studies of strong lensing thanks to surveys such as the Lenses Structure and Dynamics survey (LSD; \citealt{treu2004}), the Sloan Lens ACS Survey (SLACS; \citealt{bolton2006,koopmans2006,bolton2008a,bolton2008b,koopmans2009,auger2010a,auger2010b,shu2015,Shu2017}), the Strong Lensing Legacy Survey (SL2S; \citealt{cabanac2007,ruff2011,gavazzi2012,sonnenfeld2013a,sonnenfeld2013b,sonnenfeld2015}) and the BOSS Emission-Line Lens Survey (BELLS; \citealt{Brownstein2012}). Future surveys such as the Euclid (\citealt{lau2011}) and the Large Synoptic Survey Telescope (LSST; \citealt{Ivezic2008}), as well as the ongoing Kilo Degree Survey (KiDS; \citealt{deJong15}) and the Dark Energy Survey (DES; \citealt{DES2005}), are expected to increase the number of known strong lenses by several orders of magnitude (\citealt{petrillo2017,Metcalf2018,Treu2018}) and revolutionize  strong lensing studies.

\begin{table*} 
\begin{center}
\caption{\normalsize Main sub-grid parameters of the EAGLE simulations used in this work. Columns (left to right) are the name of the simulation, $L$: the comoving side length of the volume, $N$: the number of particles for individual type i.e. gas and DM, $\gamma_{\rm eos}$: the power-law slope of the polytropic equation of state, $n_{\rm H}^\star$: the star formation density threshold, $f_{\rm th}$: the star formation feedback efficiency, $f_{\rm th, max}$: the asymptotic maximum and $f_{\rm th, min}$: minimum values of $f_{\rm th}$, $n_{\rm H,0}$: the density-term denominator for the Reference model , $n_{\rm n}$: the Reference model density-term exponent (from equation \ref{eq:fth(Z,n)}), $C_{\rm visc}$: the sub-grid accretion disc viscosity parameter (from equation 7 in \citealt{c15}), and $\Delta T_{\rm AGN}$: the temperature increment of stochastic AGN heating. The calibrated models reproduce the GSMF at $z = 0.1$. The reference variation models adopt a single-parameter variation of the Reference simulation (varied parameters are highlighted in bold). Except for FB$\sigma$ (which uses the parameter $n_{\rm T}$), all other models have $n_{\rm Z} = 2/\ln{10}$ with the same numerical value (see equation \ref{eq:fth(T)}). For FBconst, this parameter is not applicable. This Table is partially reproduced from \citet{c15}.}
\label{tbl:simulations}
\begin{tabular}{l r r l l l l l l l l l}
\hline
\hline
Identifier & Side length  & $N$ & $\gamma_{\rm eos}$ & $n_{\rm H}^\star$ & $f_{\rm th}$-scaling & $f_{\rm th,max}$ & $f_{\rm th,min}$ & $n_{\rm H,0}$ & $n_{\rm n}$ & $C_{\rm visc}/2\pi$ & $\Delta T_{\rm AGN}$  \\
               & $L$ [cMpc]      &          &                               & [$\mathrm{cm^{-3}}$]    &                               &                        &                        & [$\mathrm{cm^{-3}}$] &            &                    & $\log_{10}$ [K]       \\
\hline
\textit{Calibrated models} \\
FBconst\        & 50  & 752      & $4/3$    & Eq.~\ref{eq:sfthreshz} & $-$             & $1.0$      & $1.0$       & $-$    & $-$          & $10^3$ & $8.5$ \\
FB$\sigma$      & 50   & 752      & $4/3$    & Eq.~\ref{eq:sfthreshz} & $\sigma_{\rm DM}^2$ & $3.0$      & $0.3$       & $-$    & $-$          & $10^2$ & $8.5$ \\
FBZ             & 50  & 752       & $4/3$    & Eq.~\ref{eq:sfthreshz} & $Z$             & $3.0$      & $0.3$       & $-$    & $-$          & $10^2$ & $8.5$ \\
Ref (FBZ$\rho$) & 50  & 752       & $4/3$    & Eq.~\ref{eq:sfthreshz} & $Z,\rho$        & $3.0$      & $0.3$       & $0.67$ & $2/\ln{10}$ & $10^0$ & $8.5$ \\
Ref-100 (FBZ$\rho$) &100  & 1504      & $4/3$    & Eq.~\ref{eq:sfthreshz} & $Z,\rho$        & $3.0$      & $0.3$       & $0.67$ & $2/\ln{10}$ & $10^0$ & $8.5$ \\
\hline
\textit{Reference-variations} \\
ViscLo          & 50  & 752       & $4/3$      & Eq.~\ref{eq:sfthreshz}    & $Z,\rho$  & $3.0$      & $0.3$            & $0.67$ & $2/\ln{10}$ & $\bm{10^2}$ & $8.5$    \\
ViscHi          & 50  & 752       & $4/3$      & Eq.~\ref{eq:sfthreshz}    & $Z,\rho$  & $3.0$      & $0.3$            & $0.67$ & $2/\ln{10}$ & $\bm{10^{-2}}$ & $8.5$   \\
AGNdT8          & 50  & 752       & $4/3$      & Eq.~\ref{eq:sfthreshz}    & $Z,\rho$  & $3.0$      & $0.3$            & $0.67$ & $2/\ln{10}$ & $10^0$ & $\bm{8.0}$      \\
AGNdT9          & 50   & 752      & $4/3$      & Eq.~\ref{eq:sfthreshz}    & $Z,\rho$  & $3.0$      & $0.3$            & $0.67$ & $2/\ln{10}$ & $10^0$ & $\bm{9.0}$      \\
NOAGN          & 50   & 752      & $4/3$      & Eq.~\ref{eq:sfthreshz}    & $Z,\rho$  & $3.0$      & $0.3$            & $0.67$ & $2/\ln{10}$ & $10^0$ & $\bm{-}$      \\
\hline
\end{tabular}
\end{center}

\end{table*}


The paper is structured as follows. In Section~\ref{simulations}, we summarize the EAGLE galaxy formation simulations and the relevant codes that are used in this paper. Section~\ref{pipeline} describes the simulation and analysis pipeline. The mass models used are described in Section~\ref{mod}. We give a brief description of the strong lensing observations in Section~\ref{observations}. In Section~\ref{comparison}, we compare mock lens samples with observations, in terms of their mass-size relations and the total matter density slopes. The implications of our results are discussed and summarized in Section \ref{discussions}. Throughout the paper, we use EAGLE simulations that assume a Chabrier stellar Initial Mass Function (IMF, \citealt{chabrier2003}) and compare these to observables derived under the same IMF assumption. The values of the cosmological parameters are $\mathrm{\Omega_\Lambda}$ = 0.693, $\mathrm{\Omega_b}$ = 0.0482519, $\mathrm{\Omega_m}$ = 0.307, ${h=H_0/(100\; {\rm km\; s^{-1} \; Mpc^{-1}})}$ = 0.6777 and ${\sigma _{8}}$ = 0.8288. These are taken from the Planck satellite data release (\citealt{planck2014}).

\section{EAGLE Simulations}\label{simulations}

Although there have been simulation studies of strong lensing focusing on the mass-size relations, the total density slope  and other observables (e.g. \citealt{remus2017,Peirani2017,xu2017}), the impacts of varying sub-grid physics (in particular baryonic feedback) on lensing statistics, their mass density slopes and stellar masses and sizes have not been studied comprehensively yet (\citealt{peirani2018}). \citet{Duffy2010} analyzed the impact of baryon physics on dark matter structure but only had low-resolution models at low redshift. 

\citet{mukherjee2018} (hereafter M18), introduced the {\tt SEAGLE} pipeline to systematically study galaxy formation via simulated strong lenses. {\tt SEAGLE} aims to investigate and possibly disentangle  galaxy formation and evolution mechanisms by comparing strong lens early-type galaxies (ETGs) from hydrodynamic simulations with those observed, analyzing them in a similar manner (although this is not always exactly possible). \\

As in M18, we make use of the Evolution and Assembly of GaLaxies and their Environments (EAGLE) simulations (\citealt{s15,c15,McAlpine2016})  -- a suite of state-of-the-art hydrodynamical simulations -- to create, model and analyze simulated strong lens-galaxies and compare them with observations. Throughout this study, we use ten selected galaxy formation scenarios (i.e.\ having different sub-grid physics prescriptions; \citealt{s15,c15}), the {\tt GLAMER} ray-tracing package (\citealt{metcalf2014,petkova2014}), and the {\tt LENSED} lens-modeling code (\citealt{tessore15a}). We preselect potential strong lenses based on their stellar masses and create projected mass maps for three different orientations. We calculate the half-mass radius from the simulated mass maps. We create mock lenses by ray tracing through the mass maps, placing an analytic \citet{sersic1968} source, at a higher redshift, having observationally motivated parameters. We ignore line-of-sight effects, which for massive ETGs is expected to be a good approximation \citep[see e.g.,][]{koopmans2006}. We use a single-orbit HST-ACS F814W noise level and PSF to mimic strong lenses found in SLACS and BELLS observations \citep{auger2010a,Bolton2012}.

Throughout this work, we also discuss possible observational systematics (e.g.\ differences in model-fitting methodologies, differences in filters/bands of the observational surveys, possible lens selection biases, etc.) as well as resolution effects in the simulations, that might affect their comparison. The main aim of this study, however, is to illustrate the effects of the sub-grid physics parametrization adopted by the EAGLE models, and the strong sensitivity of a number of strong lens observables e.g. total mass density slope, mass-size relation, and Einstein radius to the variation of the key sub-grid physics. In future work, we will analyze other properties such as the dark matter fractions and the stellar Initial Mass Function (IMF). Although we assume a Chabrier IMF in this work, the impact of assuming a different IMF (e.g. stellar mass and feedback) is partially removed during the process of calibration (see Section~\ref{cal-sim}). The impact of a changing IMF should therefore be very carefully examined and will be done in a future publication for the Reference model for which these models are available \citep[see e.g.][]{Barber2018}. A full analysis is currently not possible for the other models and well beyond the scope of this work, where we focus on the impact of galaxy-formation models.

In this section we describe the EAGLE simulations used in this study.
In Section \ref{models}, we broadly describe the types of model-variations that have been chosen and in Section \ref{subgrid}, we describe the simulation setup and the sub-grid physics recipes that are used in those model variations. Section \ref{cal-sim} describes the calibrated simulations and reference models variation are summarized in Section \ref{ref-var}. The details presented here are kept concise, yet informative, to make this paper self-contained.

\subsection{EAGLE model variations}\label{models}

The simulations explored in this paper are taken from \cite{c15} plus the 100cMpc-Reference run from \cite{s15}. \cite{c15} divided the simulations into two categories. The first comprises four simulations {\sl calibrated} to yield the $z=0.1$ galaxy stellar mass function (GSMF) and central black hole (BH) masses as a function of galaxy stellar mass. The second category comprises simulations  that each vary a single sub-grid physics parameter with respect to the Reference model but without considering whether they match the GSMF (i.e.\ they are not calibrated).
In the calibrated simulations, the models differ in terms of their adopted efficiency of feedback associated with star formation, and how this efficiency depends upon the local environment. In the Reference variation simulations, the sensitivity of the resulting galaxies to these variations are assessed. We note that similar variations have previously been done in the OWLS project \citep{schaye2010}. 
The general conclusion from previous work has been that the properties of  simulated galaxies are most sensitive to the efficiency of baryonic feedback \citep[see e.g.,][]{schaye2010, Scannapieco2012, Haas2013a, Haas2013b, vogelsberger2013}. This has motivated us to largely focus in this study on the effect of baryonic feedback on lensing observables, in particular on the total mass density profile in the inner regions of massive ETGs ($\sim$5 kpc), which was not considered during the calibration process and thus is a more reliable tracer of various formation processes.

\subsection{Subgrid physics}\label{subgrid}
Any simulation has a certain resolution limit below which the physical processes cannot be simulated via the dynamics of the particles. Similarly, the physical processes on scales smaller than the resolution of the EAGLE simulations are incorporated via analytic prescriptions. In EAGLE eleven chemical elements have been considered in the simulations. The calculations of radiative cooling and heating rates using the {\tt CLOUDY} (version 07.02) code of \cite{ferland1998}, account for variations in metallicity and for variations in the relative abundances of individual elements. The cooling rates are specified as a function of density, temperature and redshift. While implementing the cooling in EAGLE simulations, it is assumed that the optically thin gas is in a state of ionization equilibrium and is exposed to the CMB and an instantaneous, spatially uniform, temporally-evolving \citep{haardt2001} UV/X-ray background (\citealt{wiersma2009}). Stochastic star formation, as formulated by \cite{schaye2008}, has been implemented, but with the metallicity-dependent density threshold of \cite{schaye2004}. A density threshold for star formation, $n_{\rm H}^{\star}$, was imposed because star formation occurs only in cold ($T\ll 10^4 \rm K$), dense gas. Because the transition from a warm, neutral phase to a cold, molecular phase only occurs at lower densities and pressures in more metal-rich (and hence dust-rich) gas, the metallicity-dependent star formation threshold put forward by \citealt{schaye2004} (see his equations 19 and 24) was adopted:
\begin{equation} 
n_{\rm H}^\star(Z)=\min \left [ 0.1\, \rm{cm^3} \left ({\frac{Z}{0.002}}\right )^{-0.64}, 10\, \rm{cm^3} \right],
\label{eq:sfthreshz} 
\end{equation} 
where Z is the gas metallicity. 
Every star particle constitutes a stellar population with a fixed \cite{chabrier2003} IMF. The mass-to-light (M/L) ratio includes all the stellar remnants.  
The stellar evolution and mass loss implemented in EAGLE, is based on the prescription proposed in \citet{wiersma2009b}. 
The simulations adopt the stochastic thermal stellar feedback scheme of \cite{vecchia2012}, in which the temperature increment, $\bigtriangleup T_{\rm SF}$, of heated resolution elements is specified. 
The fraction of the supernova energy budget that is available for feedback determines the probability that a resolution element neighboring a young star particle is heated. This fraction is referred to as $f_{\rm th}$ \citep{vecchia2012}. According to the convention, $f_{\rm th} = 1$ equates to $1.736 \times 10^{49} \rm erg \ M^{-1}_{\odot}$, being the level of injected energy per stellar mass formed. Lastly, AGN feedback has been implemented via a single mode, where energy is injected thermally and stochastically, analogous to energy feedback from star formation.

\subsection{Calibrated simulations}\label{cal-sim}
In EAGLE model variations, the efficiency of the stellar feedback and the BH accretion were calibrated to broadly match the observed local ($z \approx$ 0) GSMF, subject to the constraint that galaxy sizes must be in agreement with observations. We explain why calibration was needed and then we briefly describe the calibrated simulations of \citet{c15}, that are also used in this paper. 
Table \ref{tbl:simulations} provides a concise overview of all the important parameters and a brief description of the four calibrated EAGLE simulations, adapted from the above-mentioned work.

\subsubsection{The necessity of calibration}
The choice of sub-grid routines and the adjustment of their parameters can result in substantial alterations of the simulation outcomes. \citet{s15} argued that the appropriate methodology for cosmological simulations is to calibrate the parameters of the uncertain sub-grid routines for feedback with a small number of key observations, in order that simulations reproduce those representative observables, and then compare properties (between simulations and observations) whose quantities that are {\sl not} considered during the calibration. The total mass density slope, examined in this paper, is one of those which was not used in calibration. The results thus obtained can reasonably be considered being a consequence of the implemented astrophysics.
On the other hand, the impact of changing the IMF  \citep[e.g.][]{Barber2018} is partly calibrated out, and will be more carefully considered in a separate paper for the Reference model.

\subsubsection{A Constant Feedback (FBconst)}\label{fbc}

This is the simplest feedback model where, independently from the local conditions, a fixed amount of energy per unit stellar mass is injected into the ISM. This fixed energy corresponds to the total energy discharged by type-II SNe ($f_{\rm th}=1$). 
The thermal stellar feedback prescription employed in EAGLE becomes inefficient at high gas densities due to resolution effects (\citealt{vecchia2012}). Model Reference (see Section \ref{ref}) compensates for this known artifact by injecting more energy at higher gas density. Because this is not done in FBconst, the stellar feedback will be less effective in high-mass galaxies (where the gas tends to have higher densities) (\citealt{c15}). 

\subsubsection{Velocity dispersion dependent feedback (FB$\sigma$)}\label{fbs}

This model prescribes stellar feedback based on the local conditions, inferred from neighboring DM particles. The efficiency, $f_{\rm th}$, is calibrated as a function of the square of the 3-dimensional velocity dispersion of the DM particles within a stellar particle's smoothing kernel at the time of its birth ($\sigma_{\rm DM}^2$). 

The prescription of $f_{\rm th}$ in its functional form, is a logistic (sigmoid) given by,
\begin{equation}
f_{\rm th} = f_{\rm th,min} + \frac{f_{\rm th,max} - f_{\rm th,min}}
{1 + \left (\frac{T_{\rm DM}}{10^5\;{\rm K}}\right )^{n_{\rm T}}}\;. 
\label{eq:fth(T)}
\end{equation}
$T_{\rm DM}$ is the temparature of the characteristic virial scale of environment of the star particle.
The parameter $n_{\rm T} > 0$ controls how rapidly $f_{\rm th}$ transitions as the dark matter `temperature' scale deviates from $10^5\;\K$. 

\subsubsection{Metallicity dependent feedback (FBZ)}\label{fbz}

This model makes the radiative losses, $f_{\rm th}$, a function of the metallicity of the ISM. Energy dissipation associated with star formation feedback are likely to be more significant when the metallicity is sufficient for cooling from metal lines to dominate over the cooling contribution from H and He. The transition of outflowing gas in the simulations is expected to occur at $Z\sim0.1\Zsun$ for a temperature range $10^5\,K < T < 10^7\,K$ \citep{wiersma2009}. This phenomenon can be numerically depicted by equation \ref{eq:fth(T)}, but only after replacing ($T_{\rm DM}$,$n_{\rm T}$,$10^5\;\rm K$) with ($Z$,$n_{\rm Z}$,$0.1\Zsun$) to obtain,
\begin{equation}
f_{\rm th} = f_{\rm th,min} + \frac{f_{\rm th,max} - f_{\rm th,min}}
{1 + \left (\frac{Z}{0.1\Zsun}\right )^{n_Z}}\;,
\label{eq:fth(Z)}
\end{equation}
where $\Zsun = 0.0127$ is the solar metallicity and $n_{\rm Z} = n_{\rm T} = 2/\ln{10}$.

\subsubsection{Reference (FB$Z\rho$)}\label{ref}

The feedback associated with FB$\sigma$ and FBZ becomes numerically inefficient in the centers of high-mass galaxies because a significant fraction of the star particles form at densities more than the resolution-dependent critical density ($\nHtc$) above which radiation loss of the feedback energy is quick (\citealt{vecchia2012}). These spurious energy losses can be partly compensated when a density dependence is introduced in the expression for $f_{\rm th}$:
\begin{equation}
f_{\rm th} = f_{\rm th,min} + \frac{f_{\rm th,max} - f_{\rm th,min}}
{1 + \left (\frac{Z}{0.1\Zsun}\right )^{n_Z} \left (\frac{n_{\rm H,birth}}{n_{{\rm H},0}}\right )^{-n_n}}\;,
\label{eq:fth(Z,n)}
\end{equation}
where $n_{\rm H,birth}$ is the gas particle's density at the time when it gets converted into a star particle. Hence, at a fixed metallicity $f_{\rm th}$ increases with density. Such a density dependence may  have a physical basis, because the star formation law and hence the feedback energy injection rate per unit volume, has a supra-linear dependence on surface density, which may result in smaller radiative losses at higher densities. 
In this work we use both the 50 and 100 cMpc boxes of the Reference model. The 100 cMpc box has a much larger number of massive galaxies for comparison to strong lens observations, whereas we use the Reference-50 boxes to compare with other model variations.

\subsection{Variations of the reference model}\label{ref-var}

\citet{s15} demonstrated that it is possible to calibrate the Reference model to reproduce the Galaxy Stellar Mass Function (GSMF) and the observed sizes (in different bands) of galaxies at $z$ = 0.1. However, a systematic study of the model's key sub-grid parameters and sensitivity of this model to the variations of sub-grid parameters are critical. In order to quantify these effects, \citet{c15} conducted a series of simulations (listed in the lower section of Table 1) for which the value of a single parameter was varied from that adopted in the Reference model. Here, we briefly summarize the five Reference model variations that are used in this work. There are five more Reference-model variations available, but those have a smaller box size (25 cMpc) that provide insufficient numbers of high-mass galaxies for comparisons to observed strong lens galaxies. 

\subsubsection{Viscosity variations (ViscLo and ViscHi)}\label{visclo}

The viscosity parameter $C_{\rm visc}$ governs two important physical processes: (a) the angular momentum scale at which gas accretion onto BHs switches from the relatively inefficient viscosity-limited regime to the Bondi-limited regime, and (b) the rate (only during the viscosity-limited regime) at which gas transits through the accretion disc \citep{Rosas-Guevara2015}.  It is important to note that in both cases (viscosity-limited and the Bondi-limited regime) are subjected to the Eddington limit. A \textit{lower} value of the viscosity parameter $C_{\rm visc}$, corresponding to a \textit{higher} sub-grid viscosity. When the sub-grid viscosity is high, an earlier onset of the dominance of AGN feedback is triggered at a larger energy injection rate during the viscosity-limited regime. The viscosity parameter could thus affect the efficiency of galaxy formation and the scale of the halo mass at the peak of the stellar fraction. Lower (higher) values for the viscosity increase (decrease) both of them. However, we note that \citet{Bower2017} showed that the transition from slow to fast black-hole growth, which leads to the quenching of star formation, occurs when the halo is sufficiently massive to make stellar feedback inefficient and depends only very weakly on $C_{\rm visc}$.

\subsubsection{Temperature variations in AGN heating (AGNdT8 and AGNdT9)}\label{agndt8}

\cite{s15} have examined the role of the AGN heating temperature in EAGLE by adopting $\Delta T_{\rm AGN}=10^{8.5}{\rm K\; and\;}10^9 {\rm K}$. They demonstrated that a higher heating temperature produces less frequent but more energetic AGN feedback episodes. They concluded it is necessary to reproduce the gas fractions and X-ray luminosities of galaxy groups. \citet{brun2014} also concluded that a higher heating temperature yields more efficient AGN feedback. We analyze two Reference-model variation simulations with $\Delta T_{\rm AGN}=10^{8} \rm K$  (AGNdT8) and $\Delta T_{\rm AGN}=10^{9} \rm K$  (AGNdT9), besides the Reference model itself which adopted $\Delta T_{\rm AGN}=10^{8.5} \rm K$. In massive galaxies, the heating events (less frequent but more energetic) are more effective at regulating star formation due to a higher heating temperature. AGNdT8 (AGNdT9) model has higher (lower) peak star fraction compared to the Reference model. The reduced efficiency of AGN feedback, when a lower heating temperature is adopted, leads to the formation of more compact galaxies, because gas can more easily accrete onto the centers of galaxies and form stars.

\subsubsection{No AGN feedback (NOAGN)}\label{noagn}

The final model that we consider has no AGN feedback and is the most extreme EAGLE model variation for massive galaxies. It appears unrealistic because the lack of AGN feedback is expected to dramatically increase the baryon concentration in the inner regions of galaxies, producing overly massive and concentrated galaxies. The reason that this variation is included, is to clearly demonstrate the effect of the absence of AGN activity. All other parameters are kept the same as in the Reference run.

\section{Creating Mock Lens Data}\label{pipeline}

Here, we explain the {\tt SEAGLE} (Simulating EAGLE LEnses) pipeline in more detail. We briefly summarize the selection criteria of the (lens) galaxies, the extraction of the galaxies from the simulations, the impact of projection on the lens galaxy convergence map (Section \ref{twin}), ray-tracing with {\tt GLAMER} to create mock lensed images (Section \ref{lenscre}), and finally the automatic process to create masks around the lensed images used in the lens modeling (Section \ref{mask}). The flow diagram shown in Figure~1 of M18 describes the {\tt SEAGLE} pipeline and the resulting data products. The reader is referred to M18 for more details on the pipeline.

\subsection{Galaxy selection and post-processing}\label{twin}

The initial down-selection of (lens) galaxies is based on the lens redshift ($z_{\rm l}$) and stellar mass ($\rm M_{\star}$) range from SLACS. \citet{auger2010a} find a broad lens redshift range of $ 0.075< z_{\rm l}  < 0.513$ and a lower limit on the total stellar mass of $\rm M_{\star} \geq 1.76 \times 10^{10} \;M_{\odot}$. The luminosities and effective radii of SLACS lens galaxies are based on a de Vaucouleurs profile fit to the galaxy brightness distribution as observed with Hubble Space Telescope (HST). We choose their I-band filter value, assuming it is closest to the bulk of the stellar mass. These are turned into stellar masses assuming either a Chabrier or Salpeter stellar IMF \citep{Salpeter1955}. We use the former in this paper to remain consistent with EAGLE. We also use a lower limit on both the line-of-sight stellar velocity dispersion ($\sigma >120 {\rm ~km\,s}^{-1}$) and the half stellar mass radius ($\mathrm{R_{50}}>1{\rm ~kpc}$) from the EAGLE snapshot catalogs to avoid blatant outliers e.g.,\ due to mergers. Table~\ref{properties} summarizes these initial selection criteria.  
%
%

We select all sub-haloes that match these selection criteria and extract all their particles from the snapshot. We do this for a single  redshift roughly in the middle of the SLACS redshift range, i.e.\ $z_{\rm l}=0.271$. We reiterate, as in M18, that the lens redshift is fixed at $z$=0.271 for all mock lenses, despite having a range of observed lens redshifts. This redshift is intermediate between that of SLACS at somewhat lower redshifts, and SL2S plus BELLS at somewhat higher redshifts. Choosing simulation boxes at different redshifts for all lenses, to account for the minor effect of evolution, is computationally not feasible. We expect the effect of evolution to be small around this redshift \citep{furlong2015a,furlong2015b} and to be smaller than the observed scatter in the inferred quantities for all galaxies. Although this neglects the effect of evolution in the simulated sample, this redshift is roughly in the middle of the bulk of the redshifts of the combined set of SLACS, BELLS and SL2S lenses. For more details on the galaxy extraction we refer to Section 3.2 of M18. 
We finally rotate the particle position vectors in several directions around the center of the lens galaxy. In the current paper, each galaxy is projected along the three simulation box axes. The particles using the same SPH kernel as used in the simulation are exported into projected surface density maps (for more specifications see \citealt{trayford2017}). For each galaxy, we separately calculate the surface  density maps for the individual particle types (DM, stars and gas), as well as their total surface density map. Stellar remnants are included in the star particles. 

\begin{table*}
\label{properties}
\begin{center}
\caption{\normalsize Summary of the simulation settings and output products.}\label{properties}
\label{properties}
\begin{tabular}{l l l l l} 
\hline
\hline
\multicolumn{5}{c}{\bf Galaxy Selection}\\
Observable & Value &Name&&Comments\\
\hline
$\rm M_{\star}$ & $\rm \geq 1.76 \times 10^{10} M_{\odot}$  &Stellar mass threshold& &Taken from \cite{auger2010a}\\
$\sigma$ & > 120 km/sec &Stellar velocity dispersion && Kept lower than SLACS\\
$\mathrm{R_{50}}$ &> 1 kpc & Projected half-mass radius&&\\
\hline
\multicolumn{5}{c}{\bf Lens Candidates}\\
&$\rm M_{\star}$ threshold &$\rm M_{\star}$ threshold& &\\
& for follow-up work & for this work& &\\

&- - - - - - - - - - - - &- - - - - - - - - - -& After 3& \\

Simulation & $\rm \geq 1.76 \times 10^{10} M_{\odot}$ & >$10^{11}\; \rm M_{\odot}$& projections & Comments \\
\hline
Reference-100cMpc&-&67& 201&100 cMpc box.\\
Reference-50 (FBZ$\rho$)&252&25 &75 &50 cMpcbox\\
FBconst&279& 22&66 &\Large{$''$}\\
FB$\sigma$&259& 22& 66&\Large{$''$}\\
FBZ&312& 19&57 &\Large{$''$}\\
ViscLo&289& 29& 87&\Large{$''$}\\
ViscHi&188& 14& 42&\Large{$''$}\\
AGNdT8&276& 27&81 &\Large{$''$}\\
AGNdT9&194& 8& 24&\Large{$''$}\\
NOAGN&312&37 & 111&\Large{$''$}\\ 
\hline
Object-properties & Value & Type&&Comments\\
\hline 

Orientation& 3&x, y, z  & &Projected surface density maps\\
Redshift & ${\rm z_{\rm l}=0.271}$& -&&Consistent with SLACS' \\
&&&& mean lens-redshift of 0.3\\
\hline
\multicolumn{5}{c}{\bf Source Properties}\\
Parameters & Value & Unit &&Comments\\
\hline
Source Type& S\'{e}rsic& -&&Consistent with SLACS lenses\\
&&&&(\citealt{newton2011})\\
Brightness&23 &apparent mag.&& \Large{$''$}\\
Size ($R_{\rm eff}$)& 0.2 &arcsec&& \Large{$''$}\\
Axis ratio ($q_s$)& 0.6&-&& \Large{$''$}\\
S\'{e}rsic Index & 1 &-&&\Large{$''$}\\
Redshift&$\rm z_{s}$=0.6&-&& \Large{$''$}\\
Position & Random &Within caustics&& Producing rings and arcs lens systems\\
&&&& consistent with SLACS\\
\hline
\multicolumn{5}{c}{\bf Instrumental Settings}\\
Parameters & Type &Value & Comments\\
\hline
PSF &  Gaussian& FWHM=0.1 arcsec& -\\
Noise & HST ACS-F814W& 2400 sec & -\\
\hline
\multicolumn{5}{c}{\bf Image Properties}\\
Map used & Properties & Value\\
\hline 
\multirow{2}{*}{Surface density}& (a) Size & 512$\times$512 pixels\\
& (b) Units&pkpc\\
\multirow{2}{*}{$\kappa$, Inv. mag. map and Lens}& (a) Size & 161$\times$161 pixels\\
& (b) Units&degrees\\
\hline
\hline
\end{tabular}
\end{center}
\end{table*}

\begin{figure*}

\rotatebox{90}{\bf \large Reference}\hspace{0.2cm}\includegraphics[width=0.7\textwidth]{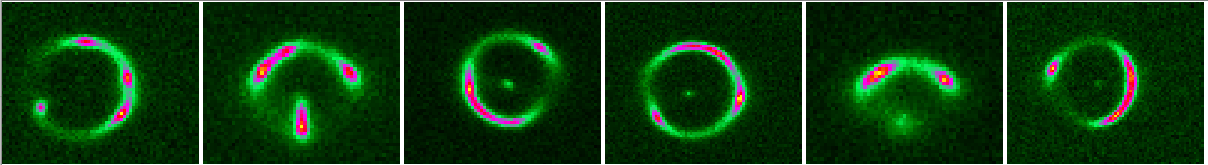}\\
\rotatebox{90}{\bf \large FBconst}\hspace{0.2cm}\includegraphics[width=0.7\textwidth]{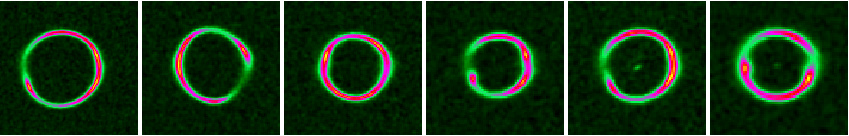}\\
\rotatebox{90}{\bf \large FB$Z$}\hspace{0.2cm}\includegraphics[width=0.7\textwidth]{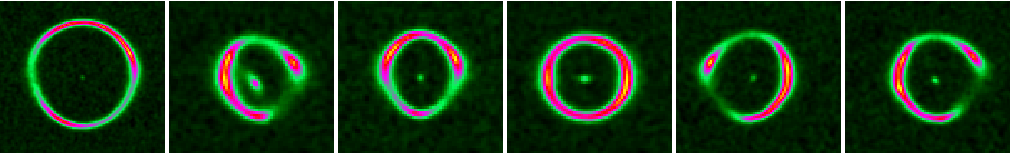}\\
\rotatebox{90}{\bf \large FB$\sigma$}\hspace{0.2cm}\includegraphics[width=0.7\textwidth]{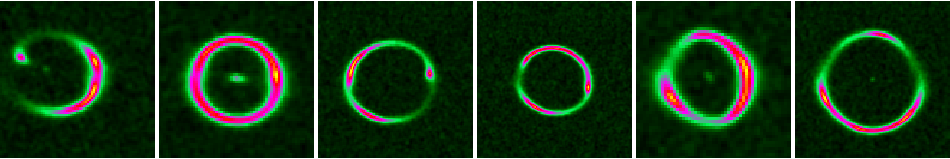}\\
\rotatebox{90}{\bf \large Visc-Lo}\hspace{0.2cm}\includegraphics[width=0.7\textwidth]{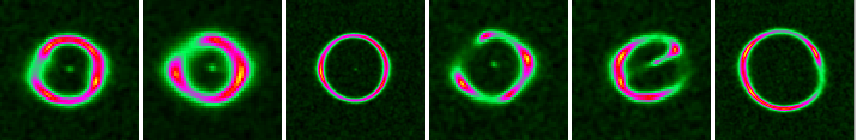}\\
\rotatebox{90}{\bf \large Visc-Hi}\hspace{0.2cm}\includegraphics[width=0.7\textwidth]{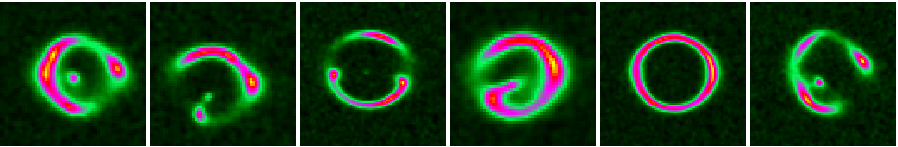}\\
\rotatebox{90}{\bf \large AGNdT8}\hspace{0.2cm}\includegraphics[width=0.7\textwidth]{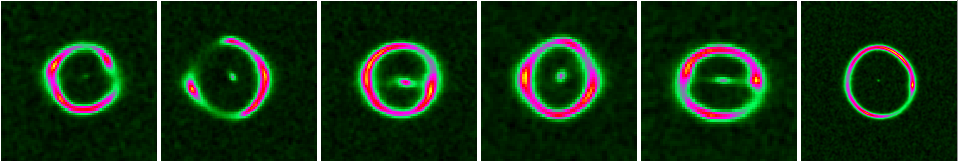}\\
\rotatebox{90}{\bf \large AGNdT9}\hspace{0.2cm}\includegraphics[width=0.7\textwidth]{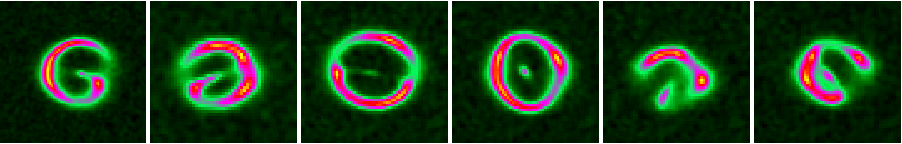}\\
\rotatebox{90}{\bf \large NOAGN}\hspace{0.2cm}\includegraphics[width=0.7\textwidth]{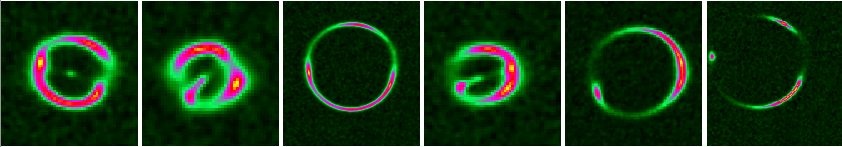}
\caption{Mosaic of a randomly selected sub-sample of six strong lenses from each of the nine EAGLE model variations ($z_{l}$ = 0.271, $z_s=0.6$). Their morphologies (for a source randomly placed inside the diamond caustic) covers that of quads, rings and arcs, and visually resemble SLACS lenses remarkably well. }
\label{mosaic1}

\end{figure*}

\subsection{Creating mock lens systems}\label{lenscre}

The surface  density maps are created in units of solar masses per pixel on a square-pixel grid of 512 $\times$ 512 (Table \ref{properties}). They form the input to the ray-tracing lensing code {\tt GLAMER} (\citealt{metcalf2014,petkova2014}). The size in proper kilo persec (pkpc) (100 pkpc) and pixel scale ($\approx$0.2 pkpc) of the grid ensure that the surface density map and the corresponding convergence map are well resolved in the inner regions of the galaxy (see \citealt{Tagore2018}), down to the simulation softening length, and are consistent with the SLACS pixel scale of 0.05 arcsec at  $z$=0.271, corresponding to $\approx$\,0.2 pkpc. 

We choose a source redshift for {\tt GLAMER} to convert these mass maps into convergence maps, by dividing the surface density maps by the critical surface density which is set by the lens and source redshifts \citep{2006glsw.conf.....M}. We choose a fixed redshift of $z_{\rm s}=0.6$, typical for SLACS lenses. Similar to the lens redshift, we choose a fixed source redshift to reduce computational overhead, although this restriction can be let go in the future. The dependence of the Einstein radius on source redshift is weak, however, increasing by $<20\%$ from $z_s=0.6$ to 1.0. Since all quantities in this work are determined inside fractions of the effective radius, the impact of the choice of the source redshift is very small.    
To describe the source, we use an elliptical S\'{e}rsic brightness profile with an index $n=1$, apparent magnitude = 23 in the HST-ACS F814W filter (AB system), an effective radius of 0.2 arcsec, a position angle $ \phi_s= 0~\deg$, and a constant axis ratio $q_s$=0.6. 
We set the parameters as such to keep close resemblance to sources found in SLACS (see  Figure  4  in  \citealt{newton2011}). As shown in M18, the choice of the source size has negligible influence on the quantities of interest in this analysis. Furthermore, in Section 4 of \citet{tessore15a}, it is shown that there is only a negligible impact on the recovered parameters when using a realistic source as opposed to using a pixellated or parametric source model.  They also show that \textsc{LENSED}  recovers the source parameters well for both an exact model (i.e.\ the truth is part of the model family) and an inexact model. Thus our constant-size analytic source model is expected to have a negligible impact on our conclusions related to the mass density slopes as is further motivated in Appendix~\ref{source-size}.

For each convergence map, the critical curves and caustics are calculated, using {\tt GLAMER}. We then randomly put the S\'{e}rsic source inside the diamond caustics of the lens to create multiple lensed images. This helps to  maximize the number of arc and ring-like systems in the simulations (this roughly mimics the large magnification bias in the observations). 
The pixel scale of the grid -- representing the lensed images -- is set to 0.05\,arcsec with the PSF and noise corresponding to an HST-ACS F814W exposure of typically 2400\,s. The final resulting images have sizes of 161 $\times$ 161 pixels with 8.0 arcsec side length. The images are exported in standard fits-file format. All parameter values are again listed  in Table \ref{properties} and  motivated mostly by the range of values inferred from SLACS lenses \citep[e.g.][]{koopmans2006,newton2011,bandara2013}.

\subsection{Mask creation}\label{mask}

To mask large areas of noisy pixels in the image and include only regions around the lensed images in the lens modeling (see Figure 4 in M18), we automatically create a mask for each lens system. The noisy lensed images are convolved with a Gaussian having FWHM of 0.25 arcsec to decrease the noise by about a factor of 5 and obtain a slightly larger footprint of the lensed images. A surface brightness threshold is set at typically 2.5--5 times below the original noise level. This threshold defines the edge of the mask, faithfully traces the lensed images below the noise, and sufficiently extends outside the lensed images to include some noise-dominated pixels in the original image (see e.g. middle panel of Figure~4 in M18). The central 7$\times$7 pixels of images (i.e.\ centered on where the lens galaxy is) are also masked, similar to what is done for real lenses. This removes any artificially bright central images that are purely the result of a too low central convergence due to the small, but still finite, size of the SPH kernel. Whereas in real lenses the central surface density in general is extremely high (i.e. leading to large gradients in the potential), thereby de-magnifying the central lensed image, in the mock lenses it leads to a too bright central image. To avoid a bias in the lens model, we mask this central region. This artificial core has however little impact on the outer images near the Einstein radius. The resulting mask is used in all subsequent modeling and only image data inside the mask are used for the lens modeling.

\section{Modeling of the Lens Sample}\label{mod}

In this section we describe the selection of the final mock lens sample (Section \ref{sample}), and the subsequent gravitational lens modeling and convergence-map fitting, i.e.\ the modeling of the surface mass density as directly obtained from the simulations (Section \ref{lensmod}). 

\subsection{The lens sample}\label{sample}

Implementing an automated recipe for the lens modeling of galaxies with stellar masses $ \rm M_{\star} < 10^{11}\; M_{\odot}$ has proven difficult due to the finite resolution effect of the particles during projection causing an artificial `core' in the inner density profile, which in turn creates prominent but artificial images in the central regions of the lenses during ray tracing. These artificial images are not observed in real lens systems and are particularly pronounced in lower-mass galaxies that are more affected by the finite resolution of the simulations. As in M18, we therefore restrict ourselves to galaxies with total stellar masses $ \rm M_{\star} > 10^{11}\; M_{\odot}$. These galaxies are far less affected by any resolution effects and still significantly overlap with the massive lensing galaxies of SLACS and SL2S. Moreover, the disc to total ratio (D/T) distributions also matches well between SLACS and EAGLE (Reference-100) and thus we should statistically select comparable ETG candidates. Of these massive galaxies, about 80\% are central galaxies (the most massive subhalo of a given halo) and about 20\% are satellites (subhalos other than the main subhalo) in the 100 cMpc box. For the 50 cMpc boxes they are mostly (>90\%) central galaxies. 
Table \ref{properties} summarizes the selection of this restricted and more massive sub-sample, used for all comparisons with observed lenses in this work. Table \ref{aaa} lists the total mass density slope (t) values and the effect of the selection bias that we introduce via a mass-weighting scheme. Table \ref{RERF} lists the average Einstein radius and several other relevant quantities of the simulated strong lenses from the different model variations of EAGLE.

\subsection{Gravitational lens modeling}\label{lensmod}

Having created the mock lens systems, we model each lens system with the open source, publicly available lens modeling code: {\tt LENSED} (\citealt{tessore15a,Bellagamba2017}). We use either an Elliptic Power Law (EPL; \citealt{tessore15b}) or a Singular Isothermal Ellipsoid (SIE; \citealt{kormann1994}) mass model, including external shear. We use the corresponding mask, noise level and PSF for each system. A total of 14 or 15 parameters are sampled using a Nested Sampling MCMC method for the SIE or EPL models, respectively.
The EPL mass model (which includes the SIE) has been utilised in several previous studies and has proven to describe very well the underlying mass model of strong gravitational lenses in various observational studies (\citealt{treu2004,koopmans2006,koopmans2009,barnabe2009,newton2011,barnabe2011}). When modeling with an SIE plus external shear, we use  the prior settings tabulated in Table 3 of M18. The SIE model's (dimensionless) surface mass density can be numerically stated as: 
\begin{equation}\label{siem}
\kappa(R)=\frac{b}{2R},
\end{equation} where $b$ equates to the measured radius of the Einstein ring  (formally only for $q$=1) and $R$ is the elliptical radius defined by:
$
R=\sqrt{q x^2 + y^2/q},
$
where $q$ is the axis ratio (minor over major axis) and $x$, $y$ are Cartesian coordinates on the image plane. The lens is allowed to vary in position angle and mass centroid as well.
We perform the lens modelling on the lenses with an EPL mass model. From \citet{tessore15b} we write the convergence as 
\begin{equation}\label{eqepl}
\kappa(R)=\frac{(2-t_L)}{2}\left( \frac{b}{R}\right)^{t_L} \; ,
\end{equation} 
where 0 < $t_L$ < 2 is the power-law surface mass density slope. The other parameters (e.g. ellipticity, position angle etc) are kept same as for the SIE model. EPL can emerge from an oblate 3D density distribution, with $\rho(r) \propto r^{-t}$, where $t=t_L+1$. Both models also include external shear parameters. 
Statistically we aim to compare the SLACS, BELLS and SL2S lenses with those from the simulated lenses via the ensemble of density slopes obtained from the EPL implemented lens-modelling technique. 

However, many of the SLACS density slopes were obtained from a joint lensing and dynamics analysis, rather than only from lensing (\citealt{koopmans2009,barnabe2009,auger2010b,barnabe2011}). We assume here that there is no significant bias between the lensing and lensing plus dynamics analyses (\citealt{Tortora14,xu2017}). A direct comparison of the model parameters with the convergence map fitting can be performed with the same model, which we do not discuss further in this work but was extensively studied in M18.
As in the creation of the mock lenses, we use a S\'{e}rsic profile to model the source. Even though some of the SLACS, BELLS and SL2S sources show irregular morphologies, our main objective is to calculate the global properties of the galaxies acting as lenses, and the exact choice of the source model does not bias the lens parameters for different (and inexact) source models (see section 4.4 of \citealt{tessore15a}). We also compare the recovered source size between SIE and EPL and found negligible difference that does not bias our results (see Appendix \ref{source-size}). In Figure \ref{fig:boat3} we also demonstrate that there is no such correlation between source-size density slope which might bias our analyses. Additional tests were carried out in M18, where we found no change in the distribution or the value of the model parameters when changing the source model parameters between lens systems (see Appendix A of M18). 
The priors used in the lens and source modeling are listed in Appendix \ref{Eapp} (see also Table 3 of M18). The priors were chosen such that the convergence of lens modeling parameters occurs faster in the Nested Sampling optimization and leads to minimal biases. We note that the priors are generally much wider than the inferred errors, hence they mostly guide the convergence rather than impact the parameter errors. The overall modelled parameters give considerably good fit to the lens and optimised residuals (for details see SEAGLE-I).



\section{Observations}\label{observations}

Here we summarize the strong lensing observational surveys that we use to compare with our results.
In Section \ref{obs} we briefly describe the observations. Section \ref{weightscheme} describes the weighting scheme that is used to compare simulated lens ensemble properties with observation. We note that in our comparison between simulated and observed lenses, we show all of the SLACS lens galaxies for visual purposes, but only quantitatively compare these galaxies with simulated galaxies for the restricted range $\rm M_{\star} > 10^{11}\; M_{\odot}$.

\subsection{SLACS, SL2S and BELLS}\label{obs}

In the SLACS survey, \citet{bolton2006} selected potential lens candidates spectroscopically from SDSS. Since then SLACS has successfully identified  more than a hundred confirmed strong lens systems, with HST follow-up. The SLACS galaxies are massive ETGs, specifically Luminous Red Galaxies (LRGs) with star-forming background sources emitting strong emission lines. The advantage of the SLACS survey is that for all lenses spectroscopic lens and source galaxy redshifts are available. The mean Einstein radius of SLACS lenses is 1.2 arcsec \citep{koopmans2006,auger2010a} with sources having a typical size of about 0.2 arcsec \citep{koopmans2006, newton2011} and typically being at $z_{\rm s} \approx 0.6$. Although it is the largest complete strong lens sample, SLACS has a relatively limited lens redshift range with the bulk of the lenses in the range of  $z_{\rm l} \approx 0.1-0.3$.

The SL2S survey was initiated to increase the number of known lenses by a different methodology than SLACS. In SL2S, \citet{cabanac2007} performed a dedicated search in the CFHTLS to find strong gravitational lenses. They focused on mostly galaxy-scale and group-scale lenses. SL2S aimed at providing a larger sample of strong lenses at higher redshift. {\tt RingFinder} (\citealt{gavazzi2014R}), an automated software was used in SL2S to find lenses by searching 170 square degrees of the sky. {\tt RingFinder} performed a search for blue arcs that are elongated tangentially and ring-like structures around red galaxies to select lens candidates. The most promising systems were followed up with HST and spectroscopy \citep{gavazzi2012}. Even though SL2S lenses combined with SLACS provided evidence of structural evolution (\citealt{ruff2011}), the SL2S sample is limited by a lack of source-galaxy redshifts for a considerable number of systems.

In BELLS, \citet{Brownstein2012} utilized the same spectroscopic methodology implemented in SLACS, to select the strong lenses, but they used Baryon Oscillation Spectroscopic Survey (BOSS; \citealt{eisenstein2011}) spectra. BELLS discovered a sample of strong galaxy-galaxy lenses, at somewhat higher redshifts, that is of comparable size and homogeneity to that of SLACS at lower redshift. BELLS is also comparable in stellar mass to the SLACS lens galaxies. Both the BELLS and SLACS samples are complete in their spectroscopic lens and source galaxy redshifts.  The lens redshifts of the three lens samples are within a similar range of 0.1–0.65, but the source redshifts cover a wide range from 0.3 to 3.5. \citet{2012ApJ...757...82B} have reported evidence for mild evolution in the mass density slope between BELLS and SLACS. We ignore this in the sample of mock lenses and compare observations with simulations only at $z= 0.271$, in between the two samples, as discussed earlier.

\subsection{Lens selection bias}\label{weightscheme}

Differences in lens-galaxy selection and follow-up can lead to differences in the population of lenses in the SLACS, BELLS and SL2S samples. For example, due to the relatively small fiber opening used in SDSS spectroscopic observations ($1.5^{\prime \prime}$ radius), the SLACS spectroscopic survey typically limits the search to lenses with an equivalent or smaller Einstein radius (although larger lenses could be found if one of the lensed images is inside the fiber and bright enough), and finite source effects play a role as well. SL2S on the other hand can select lenses directly from images and over a larger Einstein radius range, i.e.\ mass scale, typically yielding Einstein radii greater than $1^{\prime \prime}$, because they are less well resolved in ground-based images. These selection effects are hard to quantify though \citep[see e.g.,][for SLACS]{Dobler2008}. \\

Observational selection biases often hinder a proper comparison between simulations and lens surveys, strong lensing being no exception.  In this work, we assume that lens selection biases are dominated by the lens-galaxy mass and correlate sub-dominantly with the lens and source redshifts, and with the lens-galaxy mass density profile and ellipticity. This is a reasonable assumption if the lens mass models are close to isothermal (i.e.,\ the caustics are shape invariant as a function of redshift and only scale in cross-section) and the source size is small compared to the Einstein radius \citep{Dobler2008}. Massive ellipticals also do not vary strongly in their ellipticity. The observed lens sample properties are then mainly affected by the lensing cross-section (\citealt{marshall2007}), which is mass dependent, and by the magnification bias, which can be different between surveys. A precise analysis is difficult to implement and beyond the scope of this paper. We therefore take an empirical approach and only correct for the lens cross-section and we assume that the magnification bias does not correlate with galaxy mass\footnote{This holds exactly for SIE models if the source is a point source and the galaxy mass model (i.e.\ ellipticity for the SIE) does not vary with galaxy mass.}.
The square of the Einstein radius varies proportionally with the cross-section of lensing for the EPL model for a fixed ellipticity (generally also close to the SIE model). Assuming the Faber-Jackson relation (\citealt{faber1976}) and a constant mass-to-light ratio, the Einstein radius is again proportional to the stellar mass of the respective galaxy. Hence we arrive at a direct observable (i.e. the stellar mass) in both the simulations and observations. 

Motivated by the above arguments, we propose the following weighting scheme per lens:
\begin{equation}\label{w2}
\centering
W(M_\star)\equiv\left( \frac{M_\star}{\langle M_\star \rangle} \right)^{\alpha},
\end{equation}
with $\langle M_\star \rangle$ being the average stellar mass of the galaxies in the sample and $\alpha$ being the exponent of the weight function. We re-weight each simulated strong lens (which we assume to be volume limited) when comparing distributions (i.e.\ histograms) of the mass-density slopes between observed lenses from SLACS, BELLS, SL2S and simulated lenses. Hence a weight $W_i$ for simulated lens $i$ implies it counts as $W_i$ galaxies (note that the weights are non-integers). Most of the lenses are massive systems, and in general drawn from the exponential tail of the mass function. Hence re-weighting should have a limited impact on the massive end of the distribution functions, but it does strongly affect the low-mass end. We test values of $\alpha = 0.5, 1.0$, and 1.5 to show that the weighting scheme does not affect the conclusions and are only to mimic the observation selection bias of the lenses depending on their stellar mass. Other options for re-weighting the lens galaxies, to account for their lens cross-section, are using either their Einstein radii or their stellar velocity dispersions, which we have not done in this work and leave for future improvements in the analysis when we study the redshift evolution of these lenses. 
 
 
 \section{Results}\label{comparison}

In this section we compare the simulated EAGLE lenses with those from SLACS, BELLS and SL2S, in terms of their surface mass density profiles. Even though SL2S and BELLS lenses are typically at somewhat higher redshifts, we compare the simulated lenses at $z_{\rm{l}}$=0.271 assuming limited ETG evolution within the redshift range 0<$z$<1, as discussed earlier. This assumption is reasonable as it was pointed out by both \cite{sonnenfeld2013b} and \cite{koopmans2006}, that the total mass density slopes (which are close to isothermal) do not strongly evolve with time in observed ETG lenses \citep[although see][]{Bolton2012}. We compare the mass-size relation in Section \ref{mass-size1}, the total density slopes in Section \ref{density}, and the Einstein radius in Section \ref{ERC}. We compare our results with OWLS simulation in Section \ref{owls}. Table \ref{properties} summarizes the number of galaxies, lenses and projected mass maps. Tables \ref{aaa}, \ref{RERF} and \ref{slopeprop} give the effect of magnification bias (mimicked by a weighting scheme) on the total mass density slope ($t$) values, the average Einstein radii, the ratio of Einstein radius over effective radius and several other relevant quantities of the simulated strong lenses from different model variations of EAGLE.

 \subsection{The mass-size relation}\label{mass-size1}
 
 \begin{figure*}
  \centering
{\includegraphics[width=55mm]{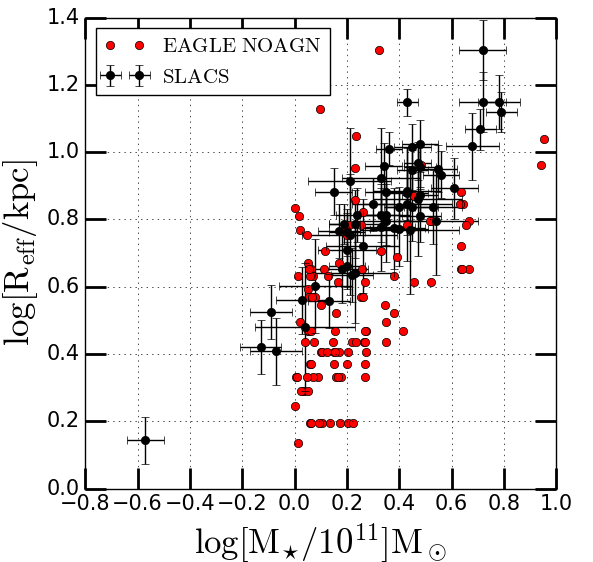}}
{\includegraphics[width=55mm]{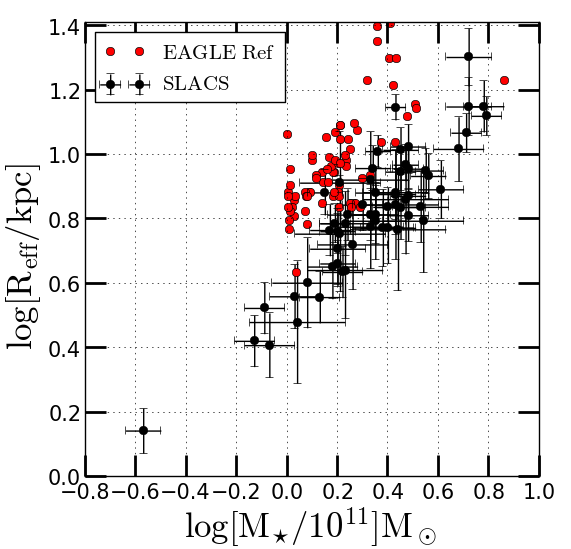}}
{\includegraphics[width=55mm]{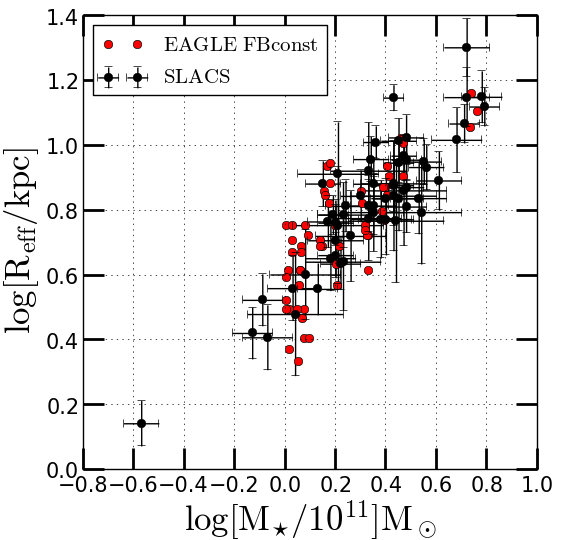}}\\
{\includegraphics[width=55mm]{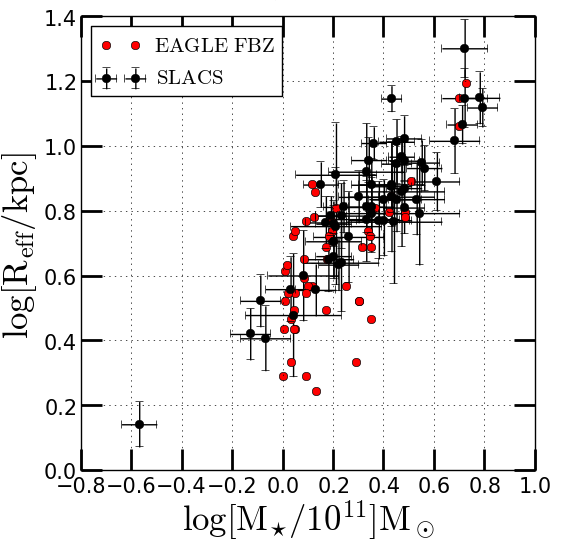}}
{\includegraphics[width=55mm]{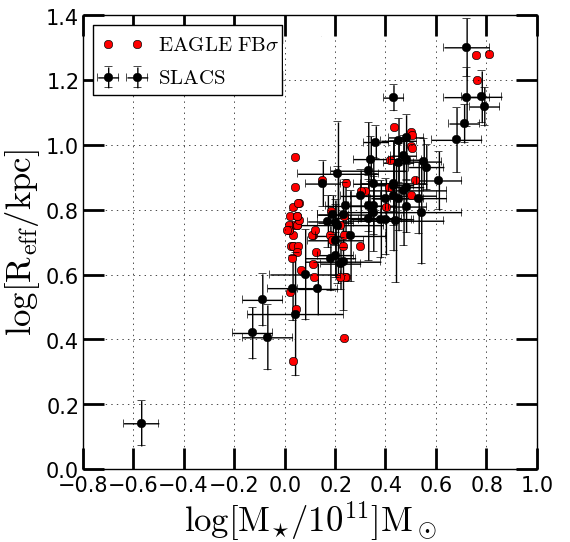}} 
{\includegraphics[width=55mm]{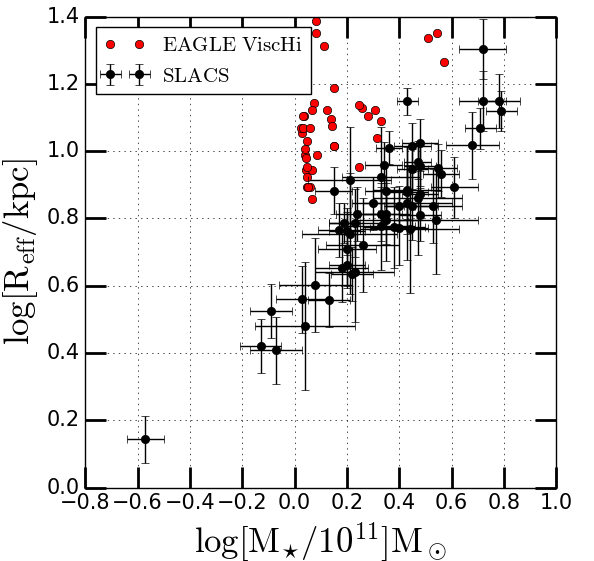}}\\
{\includegraphics[width=55mm]{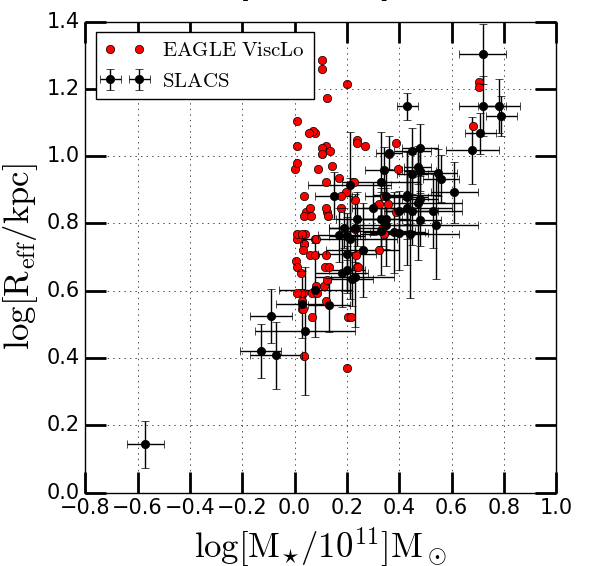}}
{\includegraphics[width=55mm]{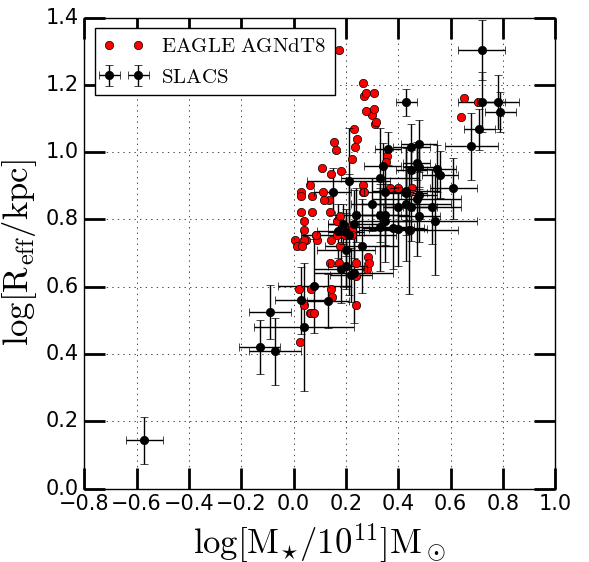}}
{\includegraphics[width=55mm]{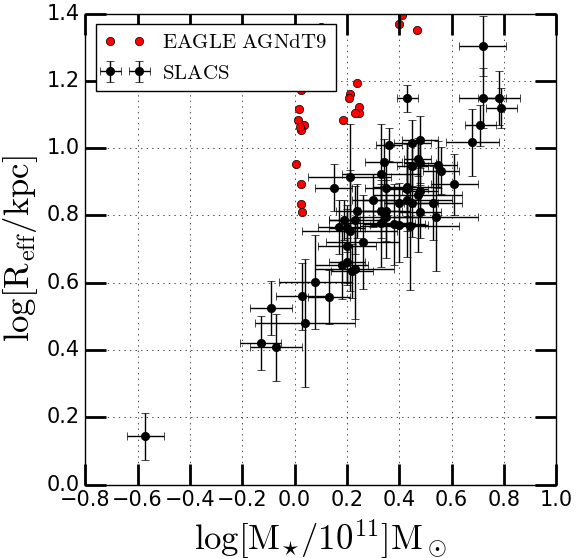}}\\
\caption{\normalsize The galaxy mass-size relation for nine EAGLE model variations from simulations with a box-size of 50 cMpc at $z_{l}$=0.271, as compared to the observed mass-size relation of SLACS lens galaxies. The stellar masses and effective radii for the observed and simulated lenses are derived using slightly different methods (fitting profiles versus inference from the simulations), but using the same stellar IMF (i.e.\ Chabrier). The simulated galaxies are only shown for stellar masses $>10^{11}$M$_\odot$, whereas for visual comparison, we show all of the SLACS lenses, although only a few of the lenses have lower stellar masses.}
\label{mass-size}
\end{figure*}
  
Observationally the stellar mass (or luminosity to be precise) of an ETG correlates with its size (e.g. \citealt{baldry2012}). Similarly, in our simulations the stellar masses of galaxies correlate with their sizes (\citealt{furlong2015b}). 
To assess whether a similar relation holds for the mock lenses at $z_{\rm l}=0.271$, we define the effective radius ($R_{\rm eff}$) as the stellar projected half-mass radius in the simulations, hence assuming a constant mass-to-light ratio. As demonstrated by \cite{remus2013,remus2017}, this might lead to a slight overestimation of the actual size of the galaxy compared to observations (e.g.\ in the case of SLACS the effective radius is derived from a de Vaucouleur fit to the galaxy brightness distribution), but we ignore this minor (<0.05 dex) effect rather than fit a profile to the projected stellar mass for all simulated galaxies. We assume a constant Chabrier IMF for both the observed and simulated galaxies. \\

Figure \ref{mass-size} shows the mass-size relations for the nine selected EAGLE model variations, overlaid on SLACS. We find that the Reference model (REF) which was calibrated on the GSMF and galaxy sizes, yields somewhat larger effective radii compared to similarly massive SLACS galaxies. On the other hand, the models FBconst, FB$\sigma$ and FBZ, which (except for FBconst) were calibrated on the GSMF but not on galaxy sizes, all have similar effective radii as SLACS, except for two outliers around the lowest stellar mass end and above the relation that have unusually large effective radii\footnote{We note that each mock lens is shown three times (once for each principle-axis orientation), as discussed earlier, and hence the number of independent outliers is very small.}. 
Due to the relatively low efficiency of stellar feedback in the FBconst, FB$\sigma$, FBZ models and the absence of AGN feedback in the NOAGN model,  stars tend to form somewhat closer to the center of the galaxy (see \citealt{c15}).
The NOAGN model, however, leads to much more compact galaxies, with some systems even straddling the resolution limit of the simulations. The galaxies from the AGN model variations (AGNdT8 and AGNdT9) both have larger effective radii than the NOAGN model. When $\Delta T=10^{8}\rm K$ (AGNdT8) about half of the galaxies are more compact in size and in good agreement with SLACS, whereas for $\Delta T=10^{9}\rm K$ (AGNdT9) hardly any galaxy matches the observations. The  higher temperature in the AGNdT9 model leads to more effective AGN feedback, keeping gas away from the center and increasing the size of the galaxy. For comparison, the Reference model assumes $\Delta T=10^{8.5}\rm K$, explaining its position halfway between AGNdT8 and AGNdT9 in mass-size relation (Figure \ref{mass-size}).
A low black hole accretion disc viscosity (ViscLo), i.e\ a high viscosity parameter ($C_{\rm visc}$), delays the onset of AGN feedback, allowing gas to settle closer to the galaxy center before star formation. The ViscHi model has the opposite effect, increasing the size of the galaxy.

Overall, we conclude that simulated galaxies from EAGLE better match the mass-size relation of SLACS lens galaxies when there is moderately low AGN activity or stellar feedback driving the galaxy formation, with only a mild impact from variations in the type of stellar feedback model. This trend is consistent with previous studies (\citealt{remus2017}; Figure 1 in \citealt{peirani2018}). Finally, we find that changes in the viscosity have a stronger impact by indirectly affecting AGN feedback.

\begin{figure}
 \includegraphics[width=\columnwidth]{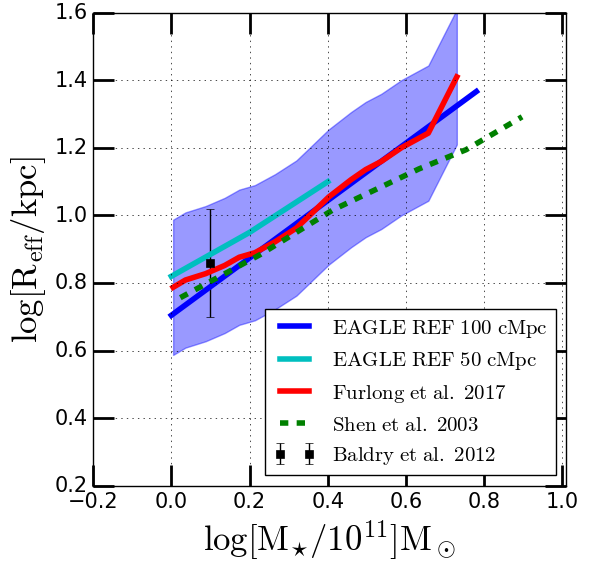}
 \caption{\normalsize Comparison of the mass-size relation obtained in this work, \citet{furlong2015b} for Reference 50 and 100 Mpc simulation box for galaxies with $M_\star$>$10^{11}\rm M_{\odot}$, \citet{shen2003} and \citet{baldry2012} are shown. The shaded region indicates the standard deviation of the spread in values.}\label{F17}
 \end{figure}
 
\subsubsection{Comparison with earlier EAGLE results}

\begin{figure*}
\includegraphics[width=0.95\columnwidth]{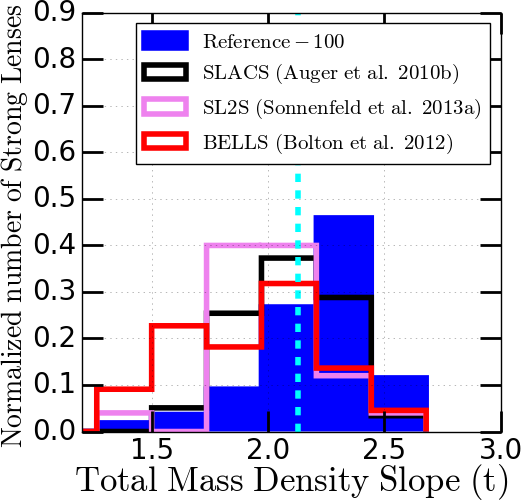}
\includegraphics[width=0.95\columnwidth]{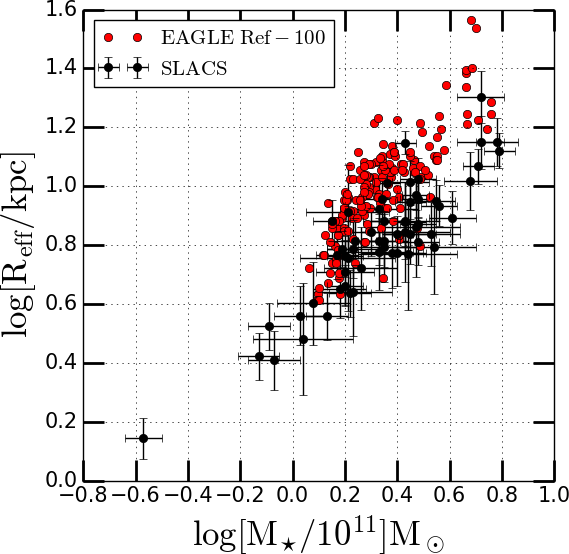}
\caption{\normalsize Left panel: Histograms of the total mass density slopes (i.e.\ $t=1-\log(\Sigma)/\log(R)$; $\Sigma(R)$ being the surface mass density of the lens galaxies) of galaxies from the EAGLE model variation Reference-100 cMpc  at $z_{l}$=0.271 (having $M_{\star}> 10^{11}{\rm M_{\odot}}$), compared to those from SLACS, BELLS and SL2S. The mean density slope from the simulations is 2.10 and the median value is 2.31. The EAGLE distributions have been obtained from lens modeling with the code {\tt LENSED}, similar in setup to the observations (see text) and have been re-weighted by a proxy of their lens cross-section to correct for the larger lens selection bias. The total mass density slopes of observations are taken from \citet{auger2010b} for SLACS, \citet{sonnenfeld2013b} for SL2S and \citet{Bolton2012} for BELLS. For SLACS and BELLS, the density slopes are derived from a combination of lensing and stellar-kinematic constraints. Right panel: The mass-size relation from the same simulation compared with SLACS. A comparative study of all the total mass density slopes (from the 50 cMpc boxes) for all other simulations is presented in Figure \ref{allslope}.}
\label{100slope}
\end{figure*}

We now compare the inferred mass-size relation with the results by \citet{s15}, \citet{c15} and \citet{furlong2015b}. This comparison is necessary to assess any selection bias within the simulations. If we are selecting an ETG population that is significantly different than the total galaxy sample analyzed in other aforementioned EAGLE works, this might invoke a bias in our lensing ETG sample and their properties. Moreover, our calculations are performed on mass maps and not directly on the cataloged particles.
\citet{s15} compared the mass-size relation of the Reference model by \citet{shen2003} and \citet{baldry2012}, and found excellent agreement. Similarly, \citet{c15} compared the mass-size relation from the calibrated models (Figure 3 therein) and found $\approx$0.2-0.3 dex difference from the Ref-50 model at the higher-mass end. This result is consistent ($\approx$0.2 dex difference) with our findings in Figure~\ref{mass-size} for our strong lensing sample. Figure~11 and 12 from \citet{c15} (3rd panel from right) show a comparison of mass-size relation of Ref-50 model variations, from which it is concluded that AGNdT9 and ViscHi models yield larger galaxy sizes compared to the AGNdT8 and ViscLo models, respectively, consistent with our findings. 
In Figure \ref{F17}, we compare the mass size relation of the Ref-100 cMpc model obtained in our analysis with that by \citet{furlong2015b}. We find excellent agreement, within 0.1 dex. We also compare with \citet{shen2003} and \citet{baldry2012} and found that our results are consistent with them. 
The mass-size trends in this paper are thus consistent with the findings of other EAGLE studies showing no bias due to our selection or methodology in calculating the sizes. As for EAGLE, part of the difference lies in the fact that the Ref-50 simulations provide larger sizes than the Ref-100 simulation at $M_\star >10^{11}\rm \Msun$, due to small number statistics for Ref-50 (see also \citealt{c15}). However, some systematic differences are still present with strong-lens galaxies tending to be more compact than non-lensing galaxies. SLACS galaxies, therefore, appear about 0.2 dex smaller in size than non-lensing galaxies of similar mass (see right panel of Figure \ref{100slope}). In paper III of SEAGLE series we will explore the systematics and compare with non-lensing ETGs from SPIDER survey \citep{Barbera2010b,Tortora14}, which we will show, have sizes that agree much better with EAGLE, and we point out the methodological differences (e.g., measurements with different bands of observations, different fitting algorithm, etc.) that could potentially bias the analysis.

\begin{figure*}
\includegraphics[width=\textwidth]{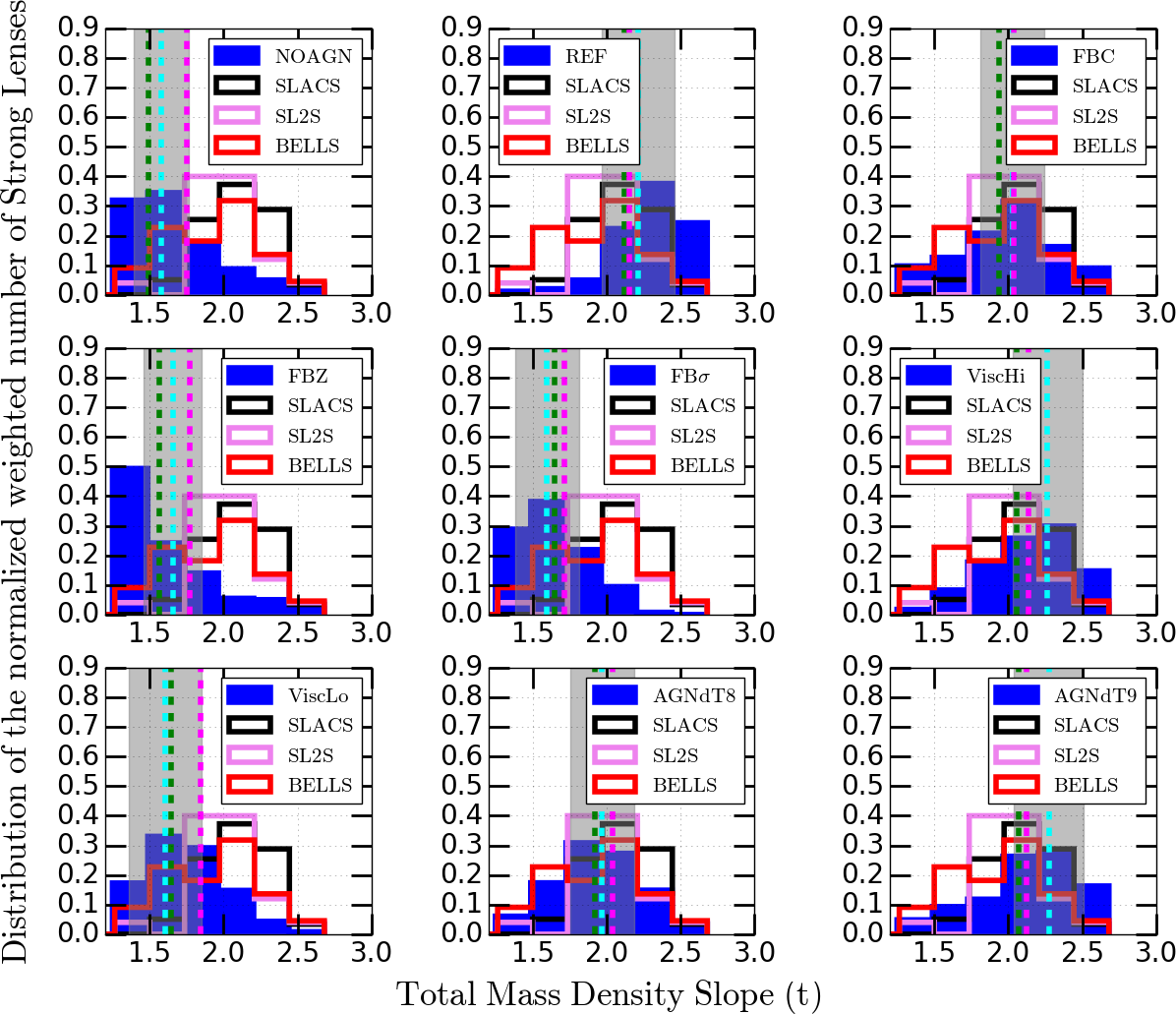}
\caption{\normalsize Histograms of the total mass density slopes (i.e.\ $t=1-d\log(\Sigma)/d\log(R)$; $\Sigma(R)$ being the surface mass density of the lens galaxies) of galaxies from EAGLE model variations (having $\rm M_{\star}> 10^{11}{\rm M_{\odot}}$) compared to those from SLACS, BELLS and SL2S. The EAGLE distributions have been obtained from lens modeling with the code {\tt LENSED}, similar in setup to the observations (see text) and have been re-weighted by a proxy for their lens cross-section to correct for the larger lens selection bias. The median values for different values of $\alpha$, see equation \ref{w2}, are shown in colored vertical dashed lines: $\alpha$=0.5 (green), $\alpha$=1.0 (cyan) and $\alpha$=1.5 (magenta). The shaded region shows the respective $\pm$ rms range centered on the median value (for $\alpha$=1.0) for each scenario. Table \ref{aaa} contains the most extreme values of $\alpha$ and their fractional difference. The total mass density slopes of observations are taken from \citet{auger2010b} for SLACS, \citet{sonnenfeld2013b} for SL2S and \citet{Bolton2012} for BELLS. For SLACS and BELLS, the density slopes are derived from a combination of lensing and stellar-kinematic constraints.}
\label{allslope}
\end{figure*}

\begin{figure*}
\centering
\includegraphics[width=0.99\textwidth]{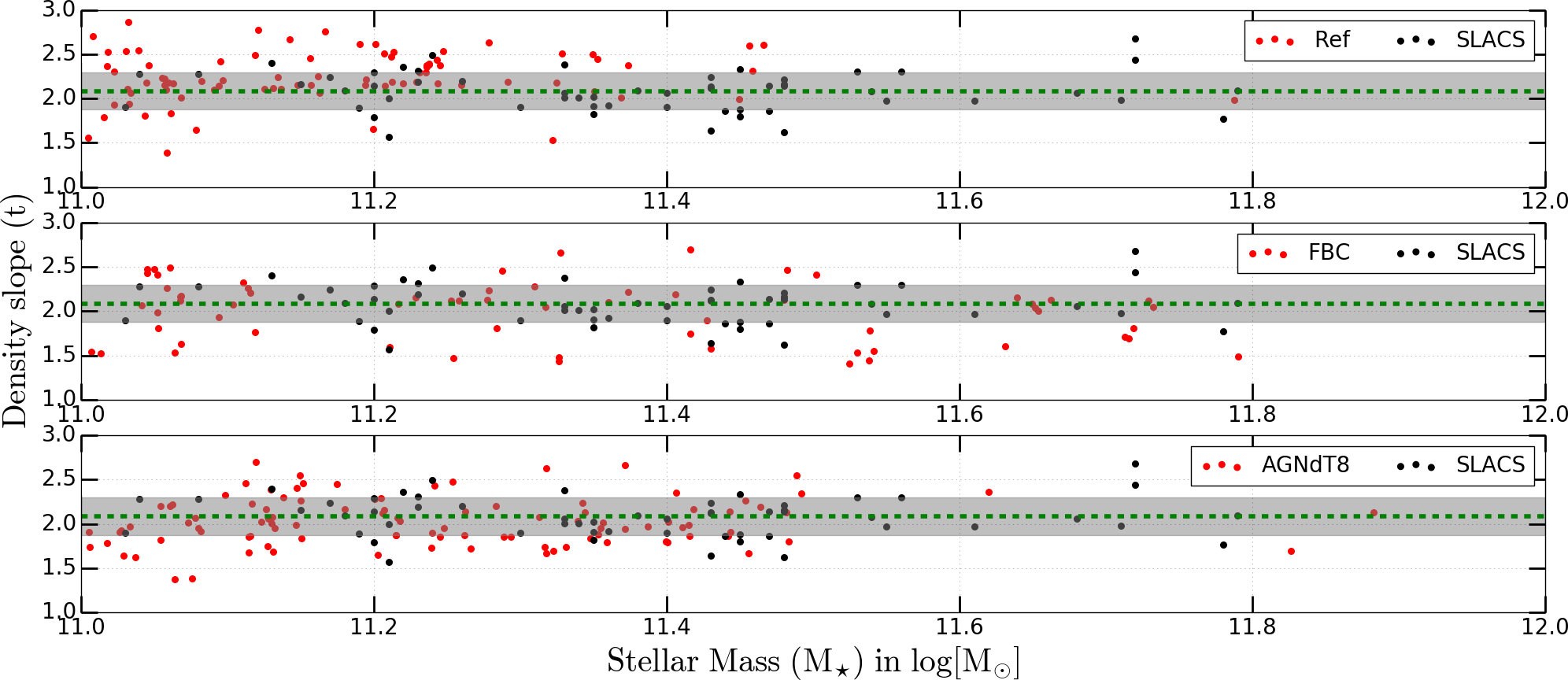}
\caption{\normalsize The total mass density slopes correlation with stellar mass from Reference, FBconst and AGNdT8 model variation of EAGLE and SLACS lenses. The mass density slope and stellar mass of SLACS lenses are obtained from \citet{auger2010b}. The dashed green line is given at SLACS mean slope at $t=2.085$ with the gray area being $\pm$10\% intrinsic scatter as obtained from \citet{koopmans2009}.}
\label{slope-SM}
\end{figure*}

 \subsection{The total mass density slope}\label{density}
 
Keeping the mass-size results discussed in the the previous section in mind, in this section we assess whether the same galaxy formation models that (visually) reproduce the mass-size relation of SLACS lens galaxies  also reproduce their mass density slopes, which is not an observable used in the calibration of the EAGLE simulations. We use the EPL surface mass density profile to model the simulated strong lenses with {\tt LENSED}, closely mimicking real lens observations (see Section \ref{mod} for details). This allows for a more unbiased and systematic comparison with strong lens observations.   

\subsubsection{Calibrated simulations}

As a first check, we confirm that the lens galaxies from the Reference-100~cMpc model show a similar distribution of density slopes as presented in M18 where the smaller 50~cMpc box was used. The latter has a much smaller number of massive galaxies. We confirm that EAGLE galaxies from the Reference model tend to have slightly steeper density slopes than SLACS, BELLS and SL2S (see left panel in Figure \ref{100slope} and also Figure 12 in M18). However, the ratio of $R_{\rm Ein}$/$R_{\rm eff}$  can play a crucial role in this respect because the lens modeling is mainly dependent on information obtained near the Einstein radius. Since the total mass density can be sensitive to the radial scale at which it is measured (\citealt{xu2017}), we will explore this aspect in Section \ref{ERC}.

In Figure~\ref{allslope} the density slopes for all EAGLE model variations are shown for the smaller 50~cMpc boxes. The FBconst model appears to yield galaxies most similar to SLACS with the total mass density profile being very close to isothermal. This can be attributed to its less efficient stellar feedback, which yields a mass profile, different than the Reference model. The FBZ and FB$\sigma$ models have more dark matter in the center of the galaxy compared to the FBconst and Reference models, leading to a shallower total density slope in their central regions. Hence, whereas the FBZ and FB$\sigma$ models visually reproduce the mass-size relation of SLACS rather well, they fail to reproduce their mass density slopes. We find the rather counterintuitive trend that when feedback efficiency increases from the FBZ, FB$\sigma$, FBconst to Reference models, the average total mass density slope steepens. We will see that variations in AGN feedback show the same trend and we will discuss the cause in the next section. In Section \ref{ERC} we will also study the correlation of the ratio of $R_{\rm Ein}$/$R_{\rm eff}$ with the total mass density slope for different model variations. 

\subsubsection{Reference-model variations}

There is a clear dependence of the total mass density slope on AGN feedback. As the stochastic temperature increment in AGN models increases from $\Delta T=10^8$K (AGNdT8) to $\Delta T=10^{8.5}$K (Reference) and $\Delta T=10^9$K (AGNdT9) the total density slope steepens. Generally, we would expect the opposite, since  stronger AGN activity (i.e.\ temperature increments) should move or keep gas particles away from the galaxy center, preventing star formation. As mentioned in \cite{brun2014}, more energetic heating events associated with a higher heating temperature, even-though less frequent, are more effective at regulating star formation in massive galaxies. \citet{c15} also pointed out that the peak galaxy formation efficiency decreases with increasing AGN temperature. The reduced efficiency of AGN feedback thus, counter-intuitively, manifests itself in a steeper total mass density slope. A similar trend is found when the viscosity parameter is increased, which impacts AGN feedback at fixed mass as discussed earlier. This trend is consistent with previous simulation studies \citep[e.g.][]{remus2017,xu2017}. In short, the AGNdT8 model with its weaker AGN feedback (compared to the Reference model) produces lensing galaxies that are closer to isothermal and in better agreement with the results from SLACS, BELLS and SL2S lens galaxies. Table \ref{prop} summarizes the mean, median and standard deviation of the density slopes for all EAGLE model variations used in this work. The evolutionary trends will be studied in details in a forthcoming paper.

\subsubsection{Correlations of slope and galaxy stellar mass}

We correlate the total mass density slope and the stellar mass of the three prominent simulation models compared in our analysis, namely, Ref-50, FBconst and AGNdT8. Figure \ref{slope-SM} shows the distribution of the density slopes calculated from lens modeling from both simulations and SLACS \citep{koopmans2009}. We find at most a very mild trend in the total mass density slope with the stellar mass, consistent with strong lensing observations of SLACS  \citep{koopmans2009}. More massive galaxies tend to have a slightly lower total density slope than less massive galaxies in all three model variations (see also \citealt{Tortora14}, where this trend, with shallower (isothermal) profile at high mass and steeper profiles at lower masses are found). However, the intrinsic scatter in the distribution in each of the model variations is too large to draw any significant conclusion, especially since the high-mass end of the distribution contains very few galaxies in the simulations. This very mild trend is also consistent with theoretical work by \citet{remus2017} and \citet{xu2017}.

\subsubsection{Dependence on weighting scheme}

We test different values of the $\alpha$ parameter in our weighting scheme to demonstrate the robustness of our results against the selection effects in the observations. In Figure \ref{allslope}, we show the variations in the median total mass density slope for three different values of $\alpha$=0.5,1.0, and 1.5.  Although, the median density slope is sensitive to the weighting scheme, relative changes are well within the spread calculated for each of the model variations. This result implies that our conclusions do not strongly depend on the observational selection bias. We note that we do not separately compensate for the magnification bias, as a function of galaxy mass,  but assume this effect is folded into the weighting scheme. The at-most mild trend of the density slope with galaxy mass, however, suggest that any re-weighting based on galaxy mass will make little difference in the conclusions. Tables \ref{aaa} and \ref{slopeprop} list the median values of the total mass density slope for different values of $\alpha$ parameter, and their relative change compared to the nominal model with $\alpha=1$. We note that we have not considered the errors on the measured slope in Figure 5. The errors on the measured slopes will slightly broaden the distributions. However, the rms error on the slopes is typically well below 0.2 (see \citealt{auger2010b}), i.e. inside our chosen bin-size, and considerably smaller than the spread in the distribution. In addition, the slope measurements from the simulations have a similar spread,  mimicking  partly this broadening effect, thus reducing its impact. The changes in galaxy-formation  processes is by far the  most prominent source differences in the distributions.  

\begin{table}
\caption{\normalsize The median values of mass density slopes, $t$, of the simulated lenses in different galaxy formation models subjected to weighting scheme with $\alpha$=0.5 and 1.5 and their respective fractional change. Table \ref{slopeprop} have the value for $\alpha$=1.0.
}\label{aaa}
\begin{center}
\begin{tabular}{l l l l}
\hline
Simulation &$\alpha$=0.5&$\alpha$=1.5& |$\Delta t$|/$t$\\
\hline
Ref-50&2.16&2.20&0.02\\
FBconst&1.98&2.08&0.05\\
FB$\sigma$&1.68&1.75&0.04\\
FBZ&1.61&1.81&0.12\\
ViscLo&1.68&1.88&0.12\\
ViscHi&2.10&2.18&0.04\\
AGNdT8&1.96&2.08&0.06\\
AGNdT9&2.11&2.17&0.03\\
NOAGN&1.61&1.79&0.11\\
\hline
\end{tabular}
 
\end{center}
\end{table}

\subsection{Einstein radius comparison and correlation with the total mass density slope}\label{ERC} 

The Einstein radius ($R_{\rm Ein}$) is a fundamental observable in strong gravitational lensing. However, to compare between strong lenses simulated from different model variations of EAGLE having a range of effective radii and strong lensing surveys having different observing strategies (e.g. SLACS, SL2S and BELLS), we need to compare the ratio of $R_{\rm Ein}$/$R_{\rm eff}$ (see \citealt{rui2018}). For SLACS, the values of $R_{\rm Ein}$/$R_{\rm eff}$ ratios populate $\approx$ 0.7 with very little scatter due to the small redshift-range for both the source and the lens (\citealt{koopmans2006,koopmans2009}). Whereas SL2S yields larger values of $R_{\rm Ein}$ with similar sized lensing ETGs as SLACS, due to the large spread in redshift-range of the lensing galaxies ($z_{\rm l}=0.2-0.8$) and the lensed sources ($z_{\rm s}=1-3.5$) \citep{sonnenfeld2013a}. In BELLS, the $R_{\rm Ein}$/$R_{\rm eff}$ values mainly range from 0.5 to 1.5 with a sharp drop below 0.5, primarily due to a wide range of the source redshift from $z_{\rm s}$=0.8 to 3.5 with mean lens redshift of $z_{\rm l}$=0.52 \citep{rui2018}. We find that our best models, FBconst and AGNdT8, are closest in their $R_{\rm Ein}$/$R_{\rm eff}$ to the mean value of SLACS. Table \ref{RERF} gives a complete overview of the mean of $R_{\rm eff}$, $R_{\rm Ein}$, the ratio $R_{\rm Ein}$/$R_{\rm eff}$ and their respective rms values for different model variations of EAGLE and observations (e.g. SLACS, SL2S and BELLS). 

\begin{figure*}
 \centering
 \includegraphics[width=0.8\textwidth]{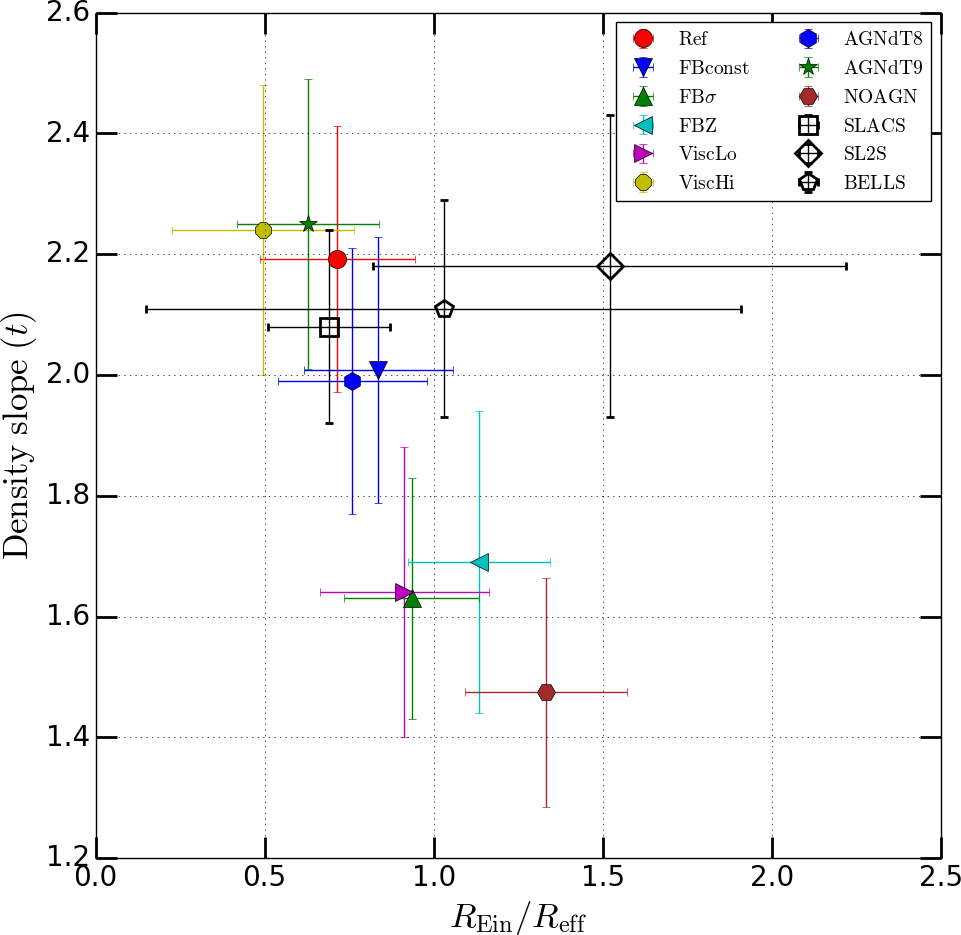}
 \caption{\normalsize Correlation of the total mass density slope ($t$) with $R_{\rm Ein}$/$R_{\rm eff}$ for nine different model variations of EAGLE and comparison with SLACS, SL2S and BELLS. The symbols used here are: FBconst (blue down-filled-triangle), FBZ (cyan left-filled-triangle), FB$\sigma$ (green up-filled-triangle), Ref (red filled-circle), AGNdT8 (blue filled-hexagon), AGNdT9 (green filled-star), ViscLo (magenta right-filled-triangle), ViscHi (orange filled-octagon), NOAGN (brown filled-hexagon), SLACS (black open-square), SL2S (black open-diamond) and BELLS (black open-pentagon).}\label{PRERF}
 \end{figure*}

Figure \ref{PRERF} shows the correlation between the average total mass-density slope ($t$) and $R_{\rm Ein}$/$R_{\rm eff}$ ratios from different model variations of EAGLE. We find that as the feedback becomes stronger, the effective radius increases (consistent with \citealt{Sales2010}). This in-turn decreases the ratio $R_{\rm Ein}$/$R_{\rm eff}$ and steepens the total density slope since $t$ is calculated at the $R_{\rm Ein}$. The larger sizes of Einstein radius for strong lenses in SL2S are primarily due to the difference in observing strategy from SLACS. SLACS and BELLS selected lens candidates from spectroscopic signatures coming from two objects at different redshifts on the same line of sight in the SDSS spectra.  The relatively small fiber used in SDSS spectroscopic observations, 1.5$^{''}$ for SLACS and 1$^{''}$ for BELLS in radius, they tend to select strong lenses with small Einstein radii. SL2S finds considerably more strong lenses with Einstein radii above 1$^{''}$, since they can be more clearly resolved in ground-based images. For similar comparison of $R_{\rm Ein}$/$R_{\rm eff}$ in SLACS, BELLS and SL2S, readers are refereed to Figure 1 in \citet{sonnenfeld2013a} and \citet{rui2018}.
 
 \subsubsection{The assumption of a power-law density profile}
 
\begin{figure}
 \centering
 \includegraphics[width=\columnwidth]{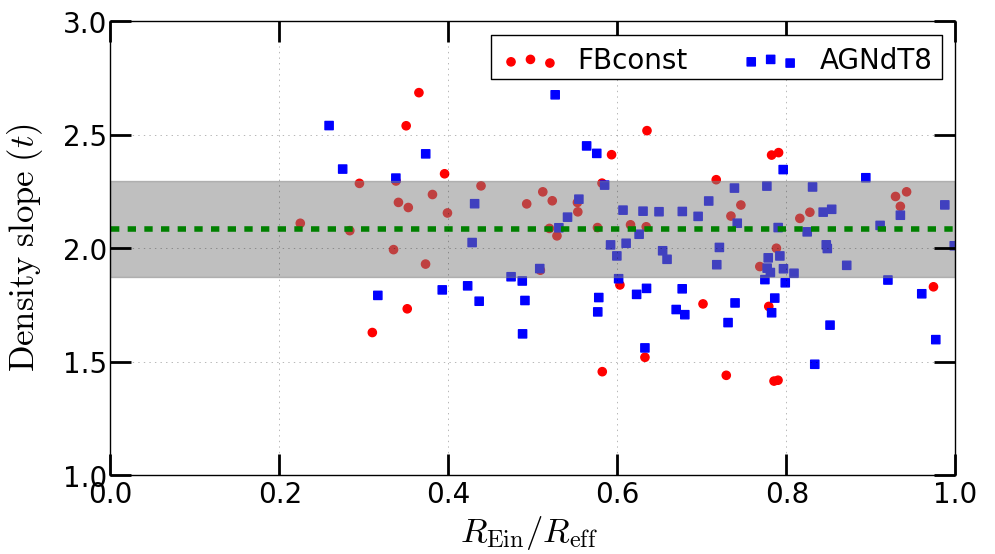}
 \caption{\normalsize Correlation of the total mass density slope ($t$) with $R_{\rm Ein}$/$R_{\rm eff}$ for individual lensing galaxies in FBconst and AGNdT8 model variations of EAGLE. The red circles are the lenses from FBconst and blue squares are from AGNdT8. The green dashed line is the mean total mass density slope of SLACS (\citealt{koopmans2009}) with $\pm$ 10\% rms (shaded region).}\label{PRERF_AM}
 \end{figure}

 \citet{koopmans2006} tested the assumption of the shape of the density-profile itself, i.e. the power-law model. If the density profiles of lens galaxies are different from a power-law, but have the same shape for each galaxy (scaled to a common scale), or, if they are different from a power-law and different between lens galaxies, the power-law assumption might give biased results. In either case, it is expected that the inferred (average) logarithmic total mass density slope inside $R_{\rm Ein}$ will change with the ratio ($R_{\rm Ein}$/$R_{\rm eff}$) for a particular model variation. In the case of the total mass density slope is a broken power-law with a change in slope inside $R_{\rm Ein}$, one expects $t$ to change depending on where the change in slope occurs with respect to $R_{\rm eff}$. Thus one is expected to find some “average” slope weighed by the luminosity and kinematic profile, varying as function of ($R_{\rm Ein}$/$R_{\rm eff}$). This is due to the dependence of $R_{\rm Ein}$ mostly on the relative distances of the lens and the source and is not a physical scale of the lens galaxy itself. \citet{koopmans2006}  found no evidence of any clear systematic correlation between $t$ and $R_{\rm Ein}$/$R_{\rm eff}$ ratio (see Figure 5 therein). Figure \ref{PRERF_AM} shows the trend in the total mass density slope and the ratio $R_{\rm Ein}$/$R_{\rm eff}$ for individual lenses. We also find no evidence of any correlation between $t$ and $R_{\rm Ein}$/$R_{\rm eff}$ ratio for both FBconst and AGNdT8 models, thus showing that our results are not biased by the power-law assumption. The small deviations of $t$ from 2.0 further support this. We conclude that our assumption of a single power-law shape for the total density profile is valid and reliable, consistent with the finding of \citet{koopmans2006}.
 
 \begin{table*}
\caption{\normalsize The mean values of effective radius, $R_{\rm eff}$, of the lensing galaxies in different galaxy formation models and their respective mean Einstein radius, $R_{\rm Ein}$. The ratio $R_{\rm Ein}/R_{\rm eff}$ gives a good estimate of the type of strong lenses simulated from EAGLE and observations.
}\label{RERF}
\begin{center}
\begin{tabular}{l c l c l l l}
\hline
Simulation &<$\log(R_{\rm eff})$>& rms &<$\log(R_{\rm Ein})$>& rms &$R_{\rm Ein}/R_{\rm eff}$& rms\\
\hline
Ref-50&0.91&	0.21&	0.65&	0.34&  0.71&0.23\\
FBconst&0.84&	0.26&	0.68&	0.35& 0.83&0.22\\
FB$\sigma$&0.82&	0.23&	0.77&	0.36& 0.94 &0.20\\
FBZ&0.72&	0.28&	0.81&	0.33& 1.13&0.21\\
ViscLo&0.83&	0.20&	0.77&	0.30& 0.93&0.25\\
ViscHi&1.08&	0.13&	0.52&	0.27& 0.46&0.27\\
AGNdT8&0.84&	0.19&	0.64&	0.28&  0.76&0.22\\
AGNdT9&1.13&	0.16&	0.71&	0.43& 0.63&0.21\\
NOAGN&0.56&	0.23&	0.75&	0.35& 1.33&0.24\\
\hdashline
SLACS& 0.86& 0.51& 0.59&0.11& 0.69&0.18\\
SL2S&0.83 &0.49 & 0.95&0.60& 1.52&0.70\\
BELLS&1.03 & 0.76& 1.05 &0.62& 1.03&0.88\\
\hline
\end{tabular}

\end{center}
\end{table*}

\subsection{Comparison with OWLS simulations}\label{owls}
In a previous study using five model variations from OWLS (\citealt{schaye2010}) and also the DM-only simulation, \citet{Duffy2010} probed the mass density slope at $z$=2 and compared the results with SLACS lenses (Figure~3 therein). They found that implementation of AGN feedback, or extremely efficient feedback from massive stars, is necessary to match the observed stellar-mass fractions in groups and clusters. However, that made the inner density profiles shallower than isothermal. They concluded that a weak or no feedback produces galaxies with an isothermal profile. This is consistent with the results in this work, where we also found that weaker feedback leads to better agreement of the total mass density slope with SLACS, BELLS and SL2S observations. However, they also conclude that other observables, such as the stellar fractions, rule out those weak feedback models (e.g. see \citealt{c15}). One way to explain this conundrum is that all the models miss something  critical, which may well be the case. Another explanation could be that the strong lenses are a biased sample of the total ETG population in a volume limited sample. 
Previously, \citet{Sales2010} explored different feedback models in OWLS (\citealt{schaye2010}) and found large variations in the abundance and structural properties of bright galaxies at $z$ = 2. They showed that models with inefficient or no feedback lead to the formation of overly massive and compact galaxies with a large fraction (upwards of 50 percent) of all available baryons (gas, stars, and stellar remnants) being retained in each halo. Increasing the efficiency of stellar or AGN feedback reduces the baryonic mass fraction fraction and increases the size of the simulated galaxies. This trend is also consistent with our findings. \\

The conclusion in \citet{Duffy2010} that NOAGN feedback produces an isothermal profile is in contradiction with our analysis. One reason could be that our analysis is carried out at a redshift of $z=0.271$, however, closer in redshift to where these lens galaxies are observed and is consistent with the results of several other simulation studies (\citealt{xu2017,remus2017}). Analysis of \citet{Duffy2010} is done at a significantly higher redshift of $z$=2. In the next section, we will discuss the possible reasons for these differences in light of potential systematics.

\subsection{Potential systematics}

There could be several effects that play a role in the comparison between observations and simulations. We describe three of these below.

\paragraph*{Evolution of the density profile: }

The inclusion of baryons results in differences in the total density profiles that depend on the efficiency of the radiative cooling and feedback.  As pointed out in \citet{remus2017} and \citet{xu2017}, there could be a significant steepening of the total mass density slope in the simulations at lower redshifts which might affect the density-slope analysis. Even though \citet{koopmans2006} have shown that there is no strong evidence for evolution in the total mass density slope in SLACS with redshift, this only holds for the redshift range of $0.1\lesssim z \lesssim 0.3$ where the bulk SLACS lenses are found. Evolution might exist as we move to higher redshifts \citep{Bolton2012, sonnenfeld2013b}. This potentially could explain the differences between this work ($z$=0.271) and the analysis in \citet{Duffy2010} which was carried out at a higher redshift ($z$=2). Moreover, the galaxies analyzed in \citet{Duffy2010} are less massive than those used in our analysis, mostly due to the significant difference in the redshifts of both the analysis. Also, for a random lens system, we measured the density profile with the lensing galaxy at three lens redshifts of $z_l$=0.101, $z_l$=0.271 and z=0.474, with the source redshift remaining at $z_s$=0.6. We found the difference in the slope parameter to be 0.02 and 0.03 respectively i.e. much below the rms error. So we assume the effects are currently not significant in our case. A similar result is also reported recently by \citet{Wang2019} where they find the density slope to remain nearly invariant after $z$=1 with a mild increase towards $z=0$. However, in our case, a full-scale redshift evolution study is beyond the scope of this work.

\paragraph*{Simulation resolution bias:}

\citet{Duffy2010} found that the resolution of the simulations can strongly affect the region where the mass density slope is measured. Their density slope measurement, however, was typically done around an Einstein radius of $\sim 3$\,kpc, only just above the resolution limit in the highest-resolution OWLS run at $z=2$. Similarly, 
\citet{schaller2015b} showed that below a radius of roughly $\sim$ 2-3 kpc, the matter density slope due to the resolution, is increasingly less reliable. This is not directly due to the softening length, but rather due to the radius enclosing a certain number of particles needed for the circular velocity to converge to within $\sim$10\% (i.e.\  the convergence radius) and the enclosed mass to within $\sim$20\%. At radii smaller than the convergence radius, the mass profile becomes increasingly less reliable and typically displays a too shallow density profiles. The impact of baryons, especially a large number of stars dominating the potential in these regions,  also becomes more uncertain.
In our work, however, we analyze galaxies at much a lower redshift and at a much higher resolution, similar to \citet{xu2017} and \citet{remus2017} (i.e.\ to Illustris and Magneticum, respectively). In these lower-redshift and higher-resolution simulations, massive galaxies have a larger Einstein radius, in the range of 3-10\,kpc, well above the resolution limit and also above the convergence radius in the simulations. We therefore expect these effects to play a minor role in the current EAGLE simulations around the Einstein radius of massive early types galaxies with $M_\star > 10^{11}$\,M${}_\odot$.

\paragraph*{Intrinsic degeneracies:}
There exists an intrinsic degeneracy between the source size and the density slope. This degeneracy is intrinsic to lensing and independent of the lens modelling technique adopted (see Appendix C \& D). The errors on the slope have a dependence on the size of the source for that particular lens-system. However, we have simulated the lenses with exactly the same source parameters (see Table 2) for uniformity. Also, in SEAGLE-I, we have tested for a small sub-sample of simulated lenses with a range of source sizes and found no evidence that a varying source size has an influence on the main modelled quantities. We do note that the error on the inferred source depends on the noise and angular resolution of the data that the referee is suggesting. In a recent published study by \citet{lyne}, it is shown that size of the mass maps could also potentially affect the image reconstruction. They also tested for several realistic slope types and finally showed that density slope and the Einstein ring measurements remained completely unaffected regardless of the change in source orientation, lensing configuration and size (see Conclusions therein). Thus when a statistical inference is drawn from the analyses presented here, it remains robust when comparing these key quantities to SLACS/BELLS/SL2S. Moreover, comparison with the fitting results coming directly from the simulations demonstrates that the recovered values for slopes via lensing are trustful to draw conclusions of their feedback mechanism. Even though, an argument maybe made that SLACS lenses have a range of source sizes, \citet{newton2011} showed along with comparison from the GEMS and HUDF samples that the values chosen in this analysis are realistic enough to simulate the lens sample regardless of that it is obtained from lenses or non-lenses. Nevertheless, as mentioned in Appendix B that the  mean source sizes obtained are 0.218, 0.217 and 0.213 for Ref, FBconst and AGNdT8 repectively which corresponds to 9\% , 8.5\% and 6.5\% mean errors respectively. Asserting that the assumption of an SIE model for the lens galaxy mass profile is the most significant source of systematic uncertainty, \citet{marshall2007} estimated the systematic errors in the source size to be 12\%. The mean errors in this analysis for key model variations are well within the expected limit for these surveys. Thus in either way the results presented in this work is robust in the scope of the parameter range that we chose to simulate the lenses.


\paragraph*{Observational biases:}

\citet{Dobler2008} found that the most significant instrumental selection effect is the finite size of the  spectroscopic fiber, which selects against large separation lenses and results in a non-monotonic dependence of the rogue line probability (defined as the probability that a given luminous red galaxy (LRG) has a rogue [O ${}_{\rm II}$] line in its spectrum) on velocity dispersion. The situation is further complicated by the effects of atmospheric seeing, which can add flux from images outside or remove flux from images inside the fiber. \citet{Dobler2008} also reported that the lensing probability has a fairly weak dependence on the size of the source (see also the appendix of M18). Hence, whereas it is clear that lens galaxies are mass-selected and biased to higher-mass galaxies, some of the most massive lenses might not have been found in SLACS due to the above-mentioned effects. These massive systems are already rare to begin with and their absence would not bias the bulk of the lens population which peaks around $M_{\star}=10^{11.35} \rm \Msun$ (\citealt{auger2010b}). As was shown by \citet{Bolton:2008p2590}, SLACS lens galaxies also appear in all observational aspects to be similar to their LRG parent population, suggesting that they are not a biased LRG sub-sample. Also, BELLS is very similar to SLACS in the type of lens galaxy, given the more heterogeneous nature of the lenses and their environments in the SL2S survey (which were morphologically and not spectroscopically selected), a lesser agreement with SL2S is maybe not entirely unexpected. Nevertheless, previous observational (e.g. \citealt{auger2010b,sonnenfeld2013b,rui2018}) and simulation analyses (\citealt{xu2016,remus2017}) these surveys have been compared among each other with the assumption that different observational selections does not hinder a fair comparison.

Moreover, as pointed out in \citet{T2014} (Table 1), strong lensing galaxies tend to be more compact than non-lensing galaxies (e.g. SPIDER sample). However, SPIDER uses K-band data and S\'{e}rsic fitted values of $R_{\rm eff}$, while SLACS uses V-band and de Vaucouleurs fit. This can give different results. But \citet{auger2010b} showed that using different fitting profiles gives negligible difference in $R_{\rm eff}$ values. Even though this is consistent with the argument that strong lensing prefers weaker feedback which in turn forms galaxies with relatively smaller sizes at fixed stellar mass compared with more efficient feedback models, it might bias correlations between galaxy properties. It could be that LRGs are a biased sub-sample of galaxies with respect to volume limited samples. We will explore this trend of galaxy sizes in light of dark matter fraction and explore possible systematics that might be causing the differences in a forthcoming work. 
\begin{table*}
\caption{\normalsize Mean, standard deviation and median values of mass density slopes inferred from lens modelling, $t$, of the simulated lenses in different galaxy formation models. The KS test results for the mass density slopes (1D) and mass-size relation (2D) compared to SLACS, BELLS and SL2S, are also listed. The p-values that exceed 0.05, and hence indicate an acceptable agreement between the simulations and observations, are shown in bold. 
}\label{prop}\label{slopeprop}
\begin{center}
\resizebox{\textwidth}{!}{
\begin{tabular}{l l l l l l l l l l l l}

\hline
\hline
\multicolumn{12}{c}{$\rm \log M_\star/M_{\odot}\; \mathrm{=11.0-12.0}$}\\
\hline
\hline
&\multicolumn{3}{c}{Mass density slope ($t$)}&\multicolumn{6}{c}{Mass density slope KS test}&\multicolumn{2}{c}{Mass-size KS test}\\
&\multicolumn{3}{c}{. . . . . . . . . . . . . . . . .}&\multicolumn{6}{c}{. . . . . . . . . . . . . . . . . . . . . . . . . . . . . . . . . . . . . . . . .}&\multicolumn{2}{c}{. . . . . . . . . . . . . . . . .}\\

Simulation&Mean & Std. & Median& \multicolumn{2}{c}{SLACS}& \multicolumn{2}{c}{SL2S}& \multicolumn{2}{c}{BELLS}&\multicolumn{2}{c}{SLACS}\\
&&&&D-value&p-value&D-value&p-value&D-value&p-value&D-value&p-value\\
\hline
Ref-100&2.09&0.26&2.24&0.26&0.53e-2&0.43&0.46e-3&0.42&0.17e-2&0.44&0.57e-2\\
Ref-50&2.19&0.25&2.20&0.35&0.15e-5&0.51&0.27e-5&0.48&0.59e-5&0.41&{\bf 0.29}\\
FBconst&2.00&0.22&2.06&0.15&{\bf 0.39}&0.36&0.005&0.17&{\bf 0.63}&0.47&{\bf 0.15}\\
FB$\sigma$&1.62&0.22&1.60&0.76&1.25e-26&0.77&4.44e-13&0.99&2.52e-19&0.48&{\bf 0.11}\\
FBZ&1.60&0.21&1.65&0.82&5.08e-27&0.84&2.23e-14&0.63&1.24e-7&0.53&0.02\\
ViscLo&1.64&0.25&1.61&0.68&1.2e-22&0.65&0.9e-10&0.46&0.001&0.52&0.002\\
ViscHi&2.09&0.23&2.24&0.17&{\bf 0.09}&0.22&{\bf 0.15}&0.21&{\bf 0.26}&0.77&1.95e-7\\
AGNdT8&1.95&0.22&2.00&0.38&{\bf 0.12}&0.36&0.003&0.21&{\bf 0.26}&0.44&{\bf 0.24}\\
AGNdT9&2.18&0.24&2.25&0.23&0.01&0.24&0.10&0.22&{\bf 0.23}&0.82&1.17e-5\\
NOAGN&1.67&0.20&1.47&0.78&5.06e-20&0.78&1.38e-11&0.51&0.11e-3&0.58&5.12e-6\\
\hline
\end{tabular}
}
\end{center}

\end{table*}

\subsection{Kolmogorov-Smirnov statistics}\label{implications}

Even though we find qualitatively and visually quite similar distributions between some of the model variations (i.e.,\ FBconst, AGNdT8) and observations, we have not quantified this (dis)agreement. 
%
%
The Kolmogorov–Smirnov (KS) test \citep{KS1933} is a nonparametric test of the equality of continuous, one-dimensional probability distributions that can be used to compare a sample with a reference probability distribution, or be used to compare two samples. The KS statistics (D-value) quantifies the maximum probability difference between the cumulative probability distribution functions of two samples. 
%
%
A KS test also yields a p-value, being the probability that two distributions are in fact drawn from the same underlying distribution and are dissimilar at the current level (D) or larger, by random chance. In this work, we use the standard 1D KS test to compare the mass density slopes and we use the 2D KS test of \citet{Peacock1983} to compare the mass-size relations. Table \ref{prop} summarizes the KS D- and p-values by comparing the results from the EAGLE model variations with those of SLACS, BELLS and SL2S, respectively. 

We indeed find that the FBconst, AGNdT8 and ViscHi models which visually appeared most consistent with the observations, also have consistently high p-values (we assume a lower limit of acceptance of $p>0.05$). 
When we combine our analysis with the p-values from the 2D KS test for the mass-size relation, we find that only the FBconst and AGNdT8 model variations remain viable. 
The Reference model, even though displaying similarity to observations of the mass-size relation from SLACS, performs poorly in the mass density slope KS test. In addition, we can clearly rule out the NOAGN, ViscLo, FBZ and FB$\sigma$ model variations based on their failure to reproduce the observed strong lens distributions in slope, mass and size. This confirms our earlier visual inspection. 


\section{Summary and Conclusions}\label{discussions}


In this paper, we have systematically explored the impact of different galaxy formation processes used in the EAGLE hydrodynamical simulations -- in particular stellar and AGN feedback -- on strong lens observables in massive early-type galaxies with $\rm M_{\star} > 10^{11} M_{\odot}$. Simulations of various mock-lens ensembles with the {\tt SEAGLE} pipeline (M18) allow us to quantify in particular the (dis)agreement between the total mass density slopes around the Einstein radius and the stellar mass-size relation between these mock lens ensembles and observations from the SLACS, BELLS and SL2S lens surveys. We compared these observables with the outcome of a range of EAGLE model variations, varying stellar \& AGN feedback and black hole accretion disc viscosity parameters (\citealt{s15,c15}). \\
%

 \noindent
We select potential strong lenses based on the stellar mass ($\rm M_\star > 10^{11}\rm \Msun$) at a redshift of $z_{\rm l}=0.271$ and create projected mass maps for three different orientations. We create mock lenses by ray tracing through the mass maps, placing an analytic \citet{sersic1968} source with observationally motivated parameters at a higher redshift ($z_{\rm s}=0.6$). We add realistic HST noise and PSF to mimic strong lenses found in observations. We calculate the projected half-mass radius for each individual mass map. We also model these lenses with an elliptic power-law model (EPL) and obtain their mass density slopes around their respective Einstein radii. Their strikingly similar visual appearance (see Figure~\ref{mosaic1}) and similar stellar mass function to SLACS, SL2S and BELLS, motivates us to compare these observed lens samples to the simulated lens systems. This allows us to compare our findings with observations and  draw the following main conclusions:\\

\noindent

\noindent (1) The stellar mass-size relation and total mass density slope of strong lens galaxies from SLACS, BELLS and SL2S agree best with EAGLE galaxy formation models that have weak or mild AGN activity or in which stellar feedback becomes inefficient at high gas densities (FBconst). In particular, the AGN model with a moderate temperature increment during active periods, $\Delta T=10^{8}\rm K$ (AGNdT8), shows excellent agreement with the observations. Models with no or high-temperature increments agree considerably less well in statistical KS tests. Similarly, the stellar-feedback model with a constant supernova energy injection per unit stellar mass into the surrounding medium  (i.e.\ FBconst) also shows excellent agreement with the observations. Our finding that more efficient feedback yields larger galaxy sizes for a fixed galaxy mass is consistent with previous work by \citet{Sales2010}, based on OWLS \citep{schaye2010}.\\

\noindent (2) Models in which the energy injection per unit stellar mass formed depends either on metalicity or local environment perform less well. Models with a high viscosity also reproduce the total mass density slopes of observed lens galaxies, but perform poorly in reproducing the mass-size relation. The EAGLE Reference model (the benchmark model) also does not perform well, most likely due to a too efficient AGN feedback model.
We note that agreement with SL2S is in general worse for all models, which we expect is due to its more heterogeneous selection (as opposed to SLACS and BELLS, they were not selected to be lenses). \\

\noindent (3) Quantitatively, we find that if the simulated lensed images are modeled using an elliptical power law (EPL) profile plus external shear, then the median total mass density slope of galaxies from the AGNdT8 and FBconst models, which have the highest $p$-values in the KS tests, are $t$=2.01 and  $t$=2.07, respectively, in good agreement with the observations of SLACS, SL2S and BELLS. Galaxies in the EAGLE Reference model, however, tend to have a steeper median total mass density slope ($t$=2.24) than observed lens galaxies (i.e.\ $t$ =2.08 for SLACS, $t$=2.11 for BELLS and $t$=2.18 for SL2S). This trend in mass density slope agrees well with the results from other independent analyses \citep[e.g.][]{remus2017, peirani2018}. \\

\noindent (4) We also assess whether in the best model variations that emerged in our analysis (FBconst and AGNdT8) and the benchmark model (Reference), $t$ correlates with stellar mass and found only a mild trend of slopes being shallower than isothermal at higher stellar mass. This is consistent with observations (\citealt{auger2010b,Tortora14}) and simulations (\citealt{remus2017,xu2017}). However, we find no evidence of correlation at any significant level between $R_{\rm Ein}$/$R_{\rm eff}$ ratios and $t$. This is consistent with \citet{koopmans2006}, \citet{auger2009}, \citet{koopmans2009} and \citet{treu2009}. Thus any selection bias based on mass should therefore not affect the conclusions. \\

\noindent (5) We also find that the mean $R_{\rm Ein}$/$R_{\rm eff}$ ratios in Reference, FBconst and AGNdT8 models are the closest to SLACS. We see a trend in the total mass density slope and $R_{\rm Ein}$/$R_{\rm eff}$ ratio where increasing the feedback efficiency, increases the $R_{\rm eff}$ thereby decreasing the value of $R_{\rm Ein}$/$R_{\rm eff}$ and steepening the total mass density slope ($t$) as in the lens modeling $t$ is calculated around $R_{\rm Ein}$. \\

\noindent Overall we conclude that more efficient feedback in massive galaxies yields steeper total mass density slopes at a radius of $\approx$ 3-10\,kpc and that strong lens galaxies appear more consistent with galaxy formation models with somewhat more limited or weaker stellar and/or AGN feedback. Our findings are consistent with the work by \citet{remus2017} and \citet{peirani2018} using different simulations. \citet{remus2017} used the Magneticum Pathfinder \citep{Hirschmann2014} and two samples, taken from zoom-in re-simulations of Oser simulations \citep{Oser2010} and Wind simulations \citep{Hirschmann2013} differing in their baryonic feedback processes. Whereas \citet{peirani2018} used two varying AGN feedback models of HORIZON-AGN simulations \citep{Peirani2017}. \\

\noindent \citet{Duffy2010} who looked at inner density slopes in the OWLS models found a similar trend, that a weaker feedback is preferred by strong lensing. However, NOAGN feedback does not produce an isothermal profile in our analysis and disagrees with \citet{Duffy2010}. These differences may be due to the fact that their mass density slope was obtained at a much higher redshift ($z$=2) and for lower-mass galaxies. Also, they did not create simulated lenses and model them with an EPL model, as done in this work, which might lead to some additional biases. We note that LRGs could have other observational selection biases and might not represent volume limited samples. Our conclusions are not biased by this trend as the evolution of $R_{\rm eff}$ is considerably small (\citealt{furlong2015b}) in EAGLE.\\

\noindent Our results prefer galaxy-formation models that have been ruled out in \citet{c15} after comparison with non-lensing observations. \citet{furlong2015b} found that the Reference model agrees well with the observed mass-size relation when compared to non-lensing galaxies. This finding is also seen in \citet{Duffy2010}, who found that weak feedback is required to match the lensing observations (consistent with our work) but also pointed out that other observables, such as the stellar fractions, rule out those weak feedback models. These seemingly opposing conclusions could be due to either differences in the precise methodologies adopted in the strong-lensing (\citealt{Duffy2010}, this work) and their non-lensing studies (\citealt{c15,furlong2015b}), or additional observational selection biases in the galaxy samples, or even from missing crucial physics.  This also might indicate that LRGs that acts as lensing galaxy, might have different formation history than the rest. A complete redshift evolution study of the total mass density slope will be addressed in a forthcoming work.  \\ 

\noindent In this work, we have demonstrated that observables of strong lens galaxies, in particular their total mass density profiles in the inner 3-10\,kpc radial range, are very sensitive to variations in the feedback in galaxy formation models. Although strong lensing analysis could have systematical difference from non-lensing analysis in the methods of the modeling. We stress again that SLACS lens galaxies are not different from the parent population of non-lens galaxies from which they were drawn (\citealt{treu2006, Bolton:2008p2590}). In paper III of the SEAGLE series we will explore the systematic errors and compare simulated lenses to non-lensing ETGs from SPIDER survey \citep{Barbera2010b,CT2012,Tortora14} and, we will show that mass-size relation of EAGLE matches very well with it. 
Whereas in this paper we have concentrated on the mass-size and mass density slopes, in forthcoming papers we will investigate the inner mass regions in more detail, focusing in particular on the effects of the dark matter distribution and the stellar IMF. 


\section*{Data Availability}
EAGLE data for all the snapshots are publicly available at \href{http://icc.dur.ac.uk/Eagle/database.php}{http://icc.dur.ac.uk/Eagle/database.php}. All the simulated lenses and other relevant quantities are available on request to SM/LVEK.  

\section*{Acknowledgements}
We thank the anonymous referee for her/his constructive suggestions to bring the paper to its present form.
SM, LVEK, CT and GV are supported through an NWO-VICI grant (project number 639.043.308). SM also acknowledges the funding from COSMICLENS: ERC-2017-ADG, Grant agreement ID: 787886. SM thanks SURFSARA network in Amsterdam. SM thanks Nicolas Tessore and Lyne Van De Vyvere for all the help and support in modeling the lenses. MS is supported by VENI grant 639.041.749. CT  also  acknowledges  funding  from  the INAF PRIN-SKA 2017 program 1.05.01.88.04. JS is supported by VICI grant 639.043.409. FB thanks the support from the grants ASI n.I/023/12/0 ``Attivit\`a relative alla fase B2/C per la missione Euclid'' and PRIN MIUR 2015 ``Cosmology and Fundamental Physics: Illuminating the Dark Universe with Euclid''. This work used the DiRAC Data Centric system at Durham University, operated by the Institute for Computational Cosmology on behalf of the STFC DiRAC HPC Facility (www.dirac.ac.uk). This equipment was funded by BIS National E-infrastructure capital grant ST/K00042X/1, STFC capital grant ST/H008519/1, and STFC DiRAC Operations grant ST/K003267/1 and Durham University. DiRAC is part of the National E-Infrastructure.  



\bibliographystyle{mnras}
\bibliography{seagle2} 




\appendix \label{app}
\section{Prior used for density slope}\label{Eapp}
Here in Table \ref{priortable}, we give the prior values used for modelling the simulated strong lenses with \textsc{LENSED}. We use a combination of uniform and Gaussian priors. In \citet{mukherjee2018}, we have explained for the motivation of priors used and also demonstrated the tests that we performed with different prior combinations for the EPL model.

\begin{table*}
\caption{\normalsize The priors used in the modeling with an EPL plus shear mass model, using {\tt LENSED}.}
\label{priortable}
\begin{center}
\begin{tabular}{l l l l l l l}
\hline
Parameter & Prior type${}^{\star \star}$ & \multicolumn{4}{c}{ Prior range}  & \multicolumn{1}{c}{Description}\\
& & $\mu $&$\sigma$& min & max & \\ 

\hline
$x_{\rm L}$      & norm &80.0 & 5.0 & -&- & Lens position: x coordinate\\                                           
$y_{\rm L}$    & norm &80.0 & 5.0 &- &- &Lens position: y coordinate\\                     
$r_{\rm L}$      & unif &- & -&5.0 & 70.0 & Einstein radius in pixel units\\
$t_{\rm L}$ & norm&1.1 &0.1&- &- & Surface mass density slope \\
$q_{\rm L}$     & unif &- &- &0.2 & 0.99& Lens axis ratio\\
${\phi}_{\rm L}$     & unif & -&- &0.0 & 180.0 &Lens position angle in degrees, wrapped around\\
${\gamma_1}_{\rm L}$     & norm& 0.0 & 0.01& -&- & Shear vector\\
${\gamma_2}_{\rm L}$     & norm &0.0 & 0.01& -& -& Shear vector\\
$x_{\rm S}$   & norm &80.0 & 30.0 & -& -& Source position: x coordinate\\                          
$y_{\rm S}$    & norm &80.0 & 30.0 &- &- & Source position: y coordinate\\                                            
$r_{\rm S}$    & unif &- & -&0.1 & 10.0 & Source size in pixel units\\
$mag_{\rm S}$  & unif &- &- &-5.0 & 0.0 & Source magnitude, adjusted with the background magnitude${}^\#$\\
$n_{\rm S}$    & norm &1.0 &  0.1 & -& -& S\'{e}rsic index \\
$q_{\rm S}$    & norm &0.5 & 0.1 & -&- & Source axis ratio \\
${\phi}_{\rm S}$   & unif & -& -&0.0 & 180.0 & Source position angle in degrees, wrapped around\\
\hline

\hline
\multicolumn{7}{l}{$\star$ All values are in pixels except $q$, $\gamma$, $t_{\rm L}$, $mag_{\rm S}$, $n_{\rm S}$, and $\phi$. $\star \star$ norm = Gaussian (with mean $\mu$ and standard dev. $\sigma$), unif = Uniform}\\
\multicolumn{7}{l}{$\#$ Source's real magnitude = Background magnitude - $mag_s$, where background magnitude is flux due to background in mag/arcsec${}^2$}
\end{tabular}

\label{priortable}
\end{center}
\end{table*}

\section{Source-size related tests}\label{source-size}
Here we present some results to demonstrate that the recovered source-sizes do not bias our conclusions. We compare the source-sizes between SIE and EPL and assess the source-size versus slope correlation in those models. These tests are in addition to those carried out in \citet{mukherjee2018}. Readers can consult the Appendix in the latter paper.

\begin{figure}
  \includegraphics[width=\linewidth]{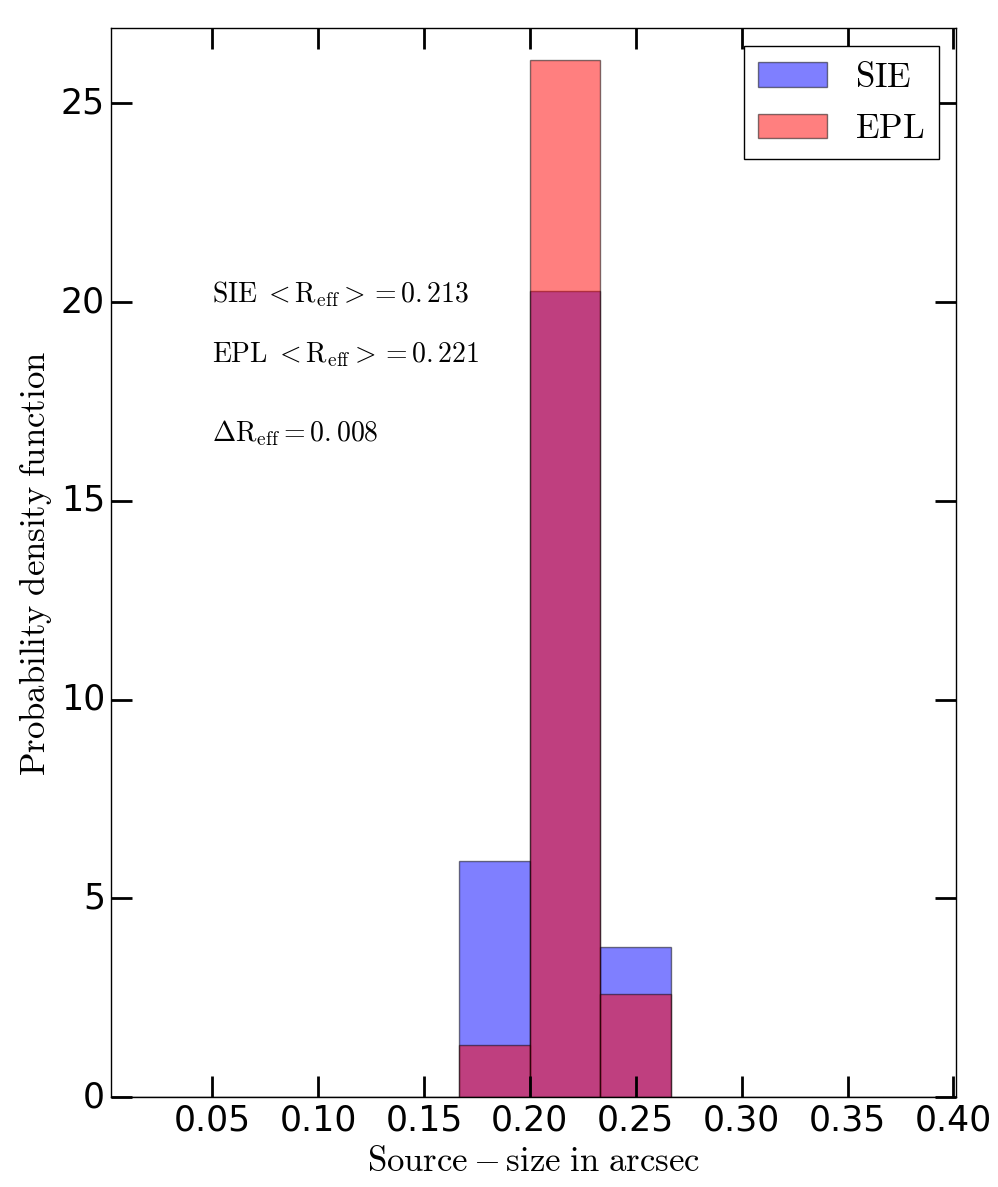}
  \caption{Source size comparison between SIE and EPL in Reference-100.}
  \label{fig:boat2}
\end{figure}

\begin{figure}
  \includegraphics[width=\linewidth]{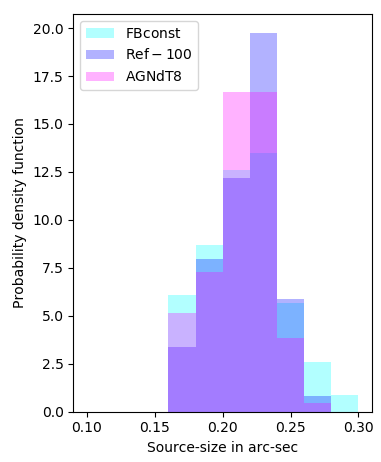}
  \caption{Source size comparison between Reference-100, FBconst and AGNdT8 sub-grid models.}
  \label{SS}
\end{figure}

\begin{figure}
  \includegraphics[width=\linewidth]{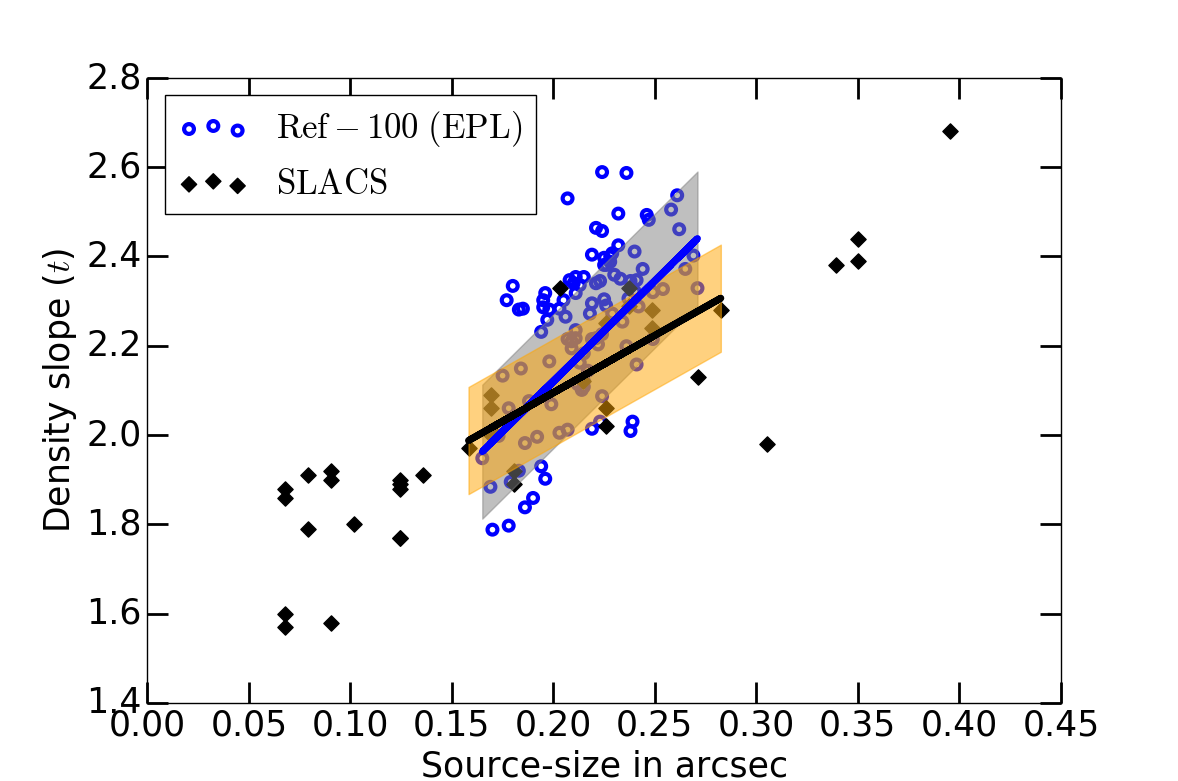}
  \caption{Source size vs mass density slope for Reference-100 simulation and SLACS. A linear function is fitted to both the Refrence-100 (EPL model) and SLACS data (blue and black line respectively). The rms error is shown by the shaded region. For visual clarity to show the difference in source-size in two models individually, see Figure B4.}
  \label{fig:boat3}
\end{figure}

\begin{figure}
  \includegraphics[width=\linewidth]{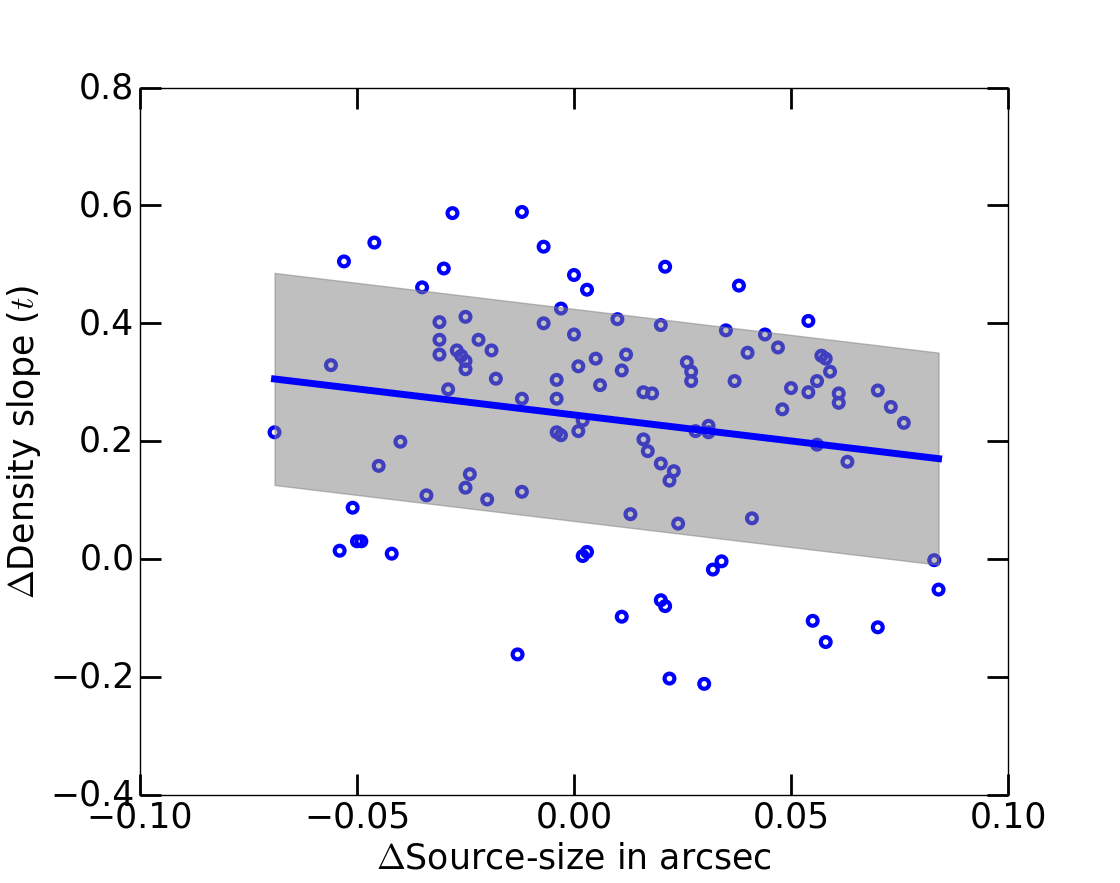}
  \caption{The difference between the EPL density slope of individual lenses from their SIE ($\equiv 2$) value for the Reference-100 model against the corresponding relative change in source size (arcsec) is shown. A very mild negative correlation with a {\tt Spearman rank} of -0.24 is found.}
  \label{fig:boat4}
\end{figure}

In Figure \ref{fig:boat2}, we present the histograms of source-size comparison between SIE and EPL for Reference-100 simulation. We show that the recovered source-sizes agree with the input ones within the error limits for both the models. Also the SIE and EPL modelling provide consistent results. The difference between source-sizes from these two different models is on average 0.008~arcsec, i.e.\ only 0.4\% of the source size. In Figure \ref{SS} We also compared the source-sizes between Reference, FBconst and AGNdT8 yielding a mean value of 0.218, 0.217 and 0.213 arcsec, respectively. Thus there is an overall perfect agreement. In Figure \ref{fig:boat3}, furthermore, we compare the source-size of SLACS and the EAGLE Reference-100 model against the density slope. We fitted a liner function to both SLACS and EPL model. We have used the same range of values for SLACS that the EPL models are covering in their sample space. We find that the EPL and SLACS slope and source-size values correlate well with each other. In Figure \ref{fig:boat4}, we also show the relative difference in source-size and EPL density slope from the SIE models. We find a mild anti-correlation having a slope of -0.09  with a {\tt Spearman rank} of -0.24. No obvious bias is found in our analyses between the EPL and SIE model values and hence we believe the conclusions to be robust.  

Finally, even if there were a small bias, such biases would occurs in real lenses as well (see \citealt{newton2011}), and hence would broaden both the observed and simulated slope distributions and not impact the inference on the formation scenarios. 

\section{Comparison with direct fitting}\label{DF}
\begin{figure}
    \centering
  \includegraphics[width=\linewidth]{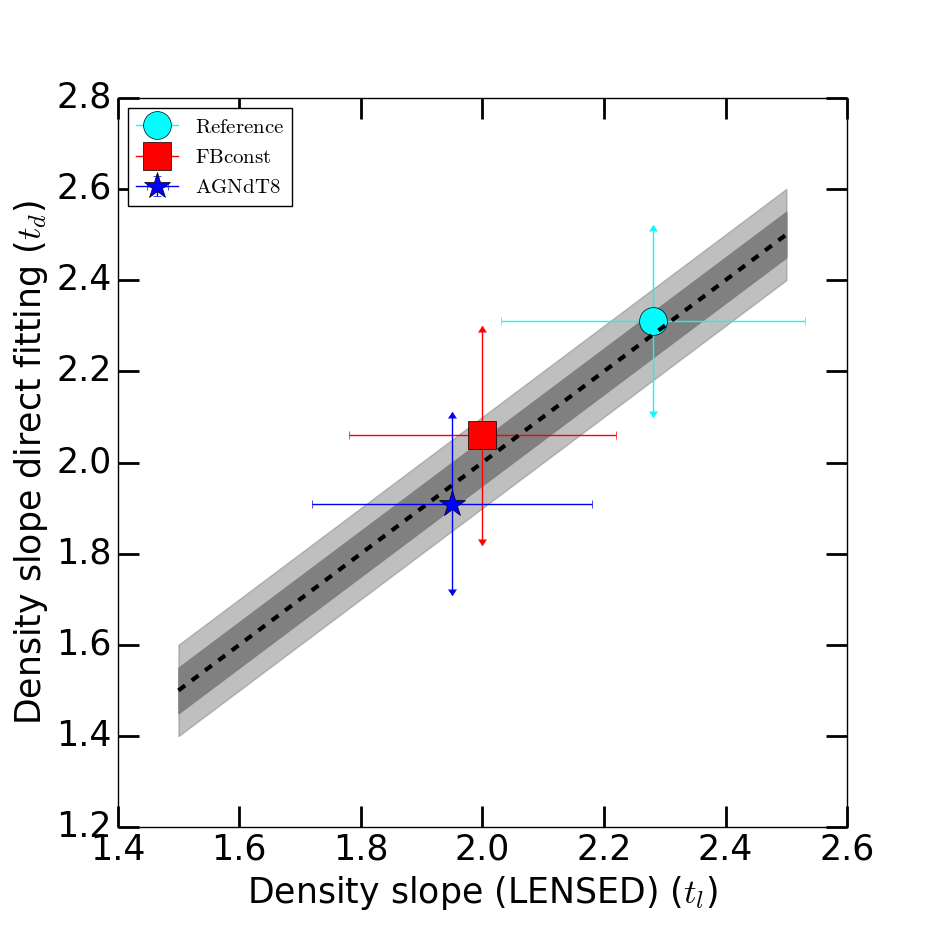}
  \caption{Comparison of mean density slopes for Reference, FBconst and AGNdT8 simulation from direct fitting and LENSED, where the error bars are the 1$\sigma$ scatter of the sample distributions. The black dashed line is the one-to-one mapping line. The dark and grey regions show the 1$\sigma$ and 2$\sigma$, respectively, where in this case $\sigma$ is the lens modeling uncertainty, i.e. 0.05.}
  \label{fig:boat5}
\end{figure}

Previously we performed this test between density slopes inferred via convergence fitting, $t_{NM}$ and LENSED, $t_{LENSED}$ in SEAGLE-I and reported (Figure. 8 therein) that there could be a difference of 10\% in Einstein radius (see also \citealt{kung2015}) and demonstrated that we find a mean ratio of 0.91 for $t_{NM}$/$t_{LENSED}$, with a standard deviation of 0.17 (Figure 9 therein). Eventhough, the lens modeling fits the density profile (more precisely that of the potential) near the lensed images, whereas the direct fit is mostly fitting the higher density regions inside the mask, we do not find any biased results from these two different methods. In Figure \ref{fig:boat5} we have shown the mean density slope comparison between Reference, FBconst and AGNdT8 models.

\section{In-exact modeling: Source parameter variation}
We have carried out several tests with a representative combination of source structures. Table \ref{inexact} summarizes the results. We use three types of complex sources: (a) two S\'{e}rsic profiles with S\'{e}rsic indices 1 and 3, (b) one S\'{e}rsic profile with index n=3, and (c) two S\'{e}rsic profiles with indices 1 and 3 plus random noise/perturbation. We fit the source with one Sersic profile. We find that the density slope, Einstein radius and ellipticity (key parameters in this work) are obtained fairly consistently. Thus we believe the conclusions in this work to be robust against modest differences between the source model and the true source structure. We do acknowledge there could be slight differences on a case by case basis, but in a statistical sense the inferences from this analysis are not affected by the source structure, and our modelling behaves similarly between real and simulated lenses. We also tested if the impact of source size have any selective effects on model variations. Table \ref{testonmodels} summarizes the results. We chose a random lensing galaxy from each model variation in the stellar mass range $M_* \sim 10^{11.0-11.1}M_\odot$. We use different source sizes other than the one used in this analysis. However, we find consistent result for mass density slope with variations that are appreciably lower than the average r.m.s. error that ranges from 0.1--0.2.

 \begin{table}
  \begin{center}
\caption{Comparison of the modeled density slopes and key parameters using different source structures for a typical EAGLE lens. The source redshift is 0.6 and lens redshift is 0.271. The remaining settings are the same as mentioned in the modeling section of the paper.} 
\begin{tabular}{lllll}
\\
\hline \hline\\
\multicolumn{3}{c}{Input S\'{e}rsic source parameters}\\
\multicolumn{3}{c}{$R_{\rm Sersic}$=0.1 arcsec, $n_{\rm Sersic}$= 1 and 3}\\

\hline
\multicolumn{3}{c}{\textbf{Modelled output}}\\

\textbf{Parameters}                  & \textbf{SIE}     & \textbf{EPL}    \\
\hline \hline\\
${\rm R}_{\rm Ein} \; (\rm arcsec)$                 & 2.62   & 2.62      \\ \hline
${\rm Density \; Slope\; (t)}$              & $\equiv$2.00   & 2.049      \\ \hline
${\rm Ellipticity}_{\rm Lens}$      & 0.158   & 0.158      \\ \hline
${\rm PA}_{\rm Lens}$               & 78.17 & 77.98  \\ \hline
${\rm Source\;} {} (n_{\rm Sersic})$             & 1.637   & 1.617     \\ \hline
${\rm Source\; }R_{\rm Sersic}\; (\rm arcsec)$              & 0.097   & 0.118    \\ \hline
$ \chi^2 $ & 0.961   & 0.959   \\ \hline \hline \\
\multicolumn{3}{c}{Input S\'{e}rsic source parameters}\\
\multicolumn{3}{c}{$R_{\rm Sersic}$=0.1 arcsec, $n_{\rm Sersic}$= 3}\\
\hline
\multicolumn{3}{c}{\textbf{Modelled output}}\\

\textbf{Parameters}                  & \textbf{SIE}     & \textbf{EPL}    \\
\hline \hline\\
${\rm R}_{\rm Ein} \; (\rm arcsec)$                 & 2.62   & 2.63      \\ \hline
${\rm Density \; Slope\; (t)}$              & $\equiv$2.00   & 2.045      \\ \hline
${\rm Ellipticity}_{\rm Lens}$      & 0.157   & 0.158      \\ \hline
${\rm PA}_{\rm Lens}$               & 78.04 & 78.63  \\ \hline
${\rm Source\;} {} (n_{\rm Sersic})$             & 2.754   & 2.771     \\ \hline
${\rm Source\; }R_{\rm Sersic}\; (\rm arcsec)$              & 0.107   & 0.095    \\ \hline
$ \chi^2 $ & 0.925   & 0.916   \\ \hline \hline \\
\multicolumn{3}{c}{Input S\'{e}rsic source parameters}\\
\multicolumn{3}{c}{$R_{\rm Sersic}$=0.1 arcsec, $n_{\rm Sersic}$= 1 and 3, random noise }\\
\hline
\multicolumn{3}{c}{\textbf{Modelled output}}\\

\textbf{Parameters}                  & \textbf{SIE}     & \textbf{EPL}    \\
\hline \hline\\
${\rm R}_{\rm Ein} \; (\rm arcsec)$                 & 2.61   & 2.61      \\ \hline
${\rm Density \; Slope\; (t)}$              & $\equiv$2.00   & 1.988      \\ \hline
${\rm Ellipticity}_{\rm Lens}$      & 0.158   & 0.156      \\ \hline
${\rm PA}_{\rm Lens}$               & 78.95 & 78.63  \\ \hline
${\rm Source\;} {} (n_{\rm Sersic})$             & 1.672   & 1.634     \\ \hline
${\rm Source\; }R_{\rm Sersic}\; (\rm arcsec)$              & 0.105   & 0.102    \\ \hline
$ \chi^2 $ & 0.922   & 0.914   \\ \hline \hline

\end{tabular}
\label{inexact}
\end{center}
\end{table}


\begin{table*}
\begin{center}
\caption{Comparison of the modeled density slopes from different model variations of EAGLE than Reference with different source sizes from that used in the main analysis.} 
\begin{tabular}{lllllllll}
\\
Source sizes${}^{\dagger}$ & FBconst & FBZ & FB$\sigma$ & AGNdT8 & AGNdT9 & ViscLo & ViscHi & NOAGN   \\ 
- - - - - - - -  & - - - - - -  & - - - -  & - - - - & - - - - - -  & - - - - - - - & - - - - - & - - - - - & - - - - -   \\ 
(arcsec) & \multicolumn{8}{c}{Mass density slope ($t$) } \\\hline
\hline
0.05 & 2.00 & 1.66 & 1.62 & 1.99 & 2.21 & 1.62 & 2.18 & 1.49 \\ \hline
0.1 & 2.01 & 1.66 & 1.62 & 1.97 & 2.19 & 1.60 & 2.16 & 1.55 \\ \hline
0.3 & 2.01 & 1.65 & 1.60 & 1.98 &2.19  & 1.61 & 2.16 & 1.57 \\ \hline
0.4 & 2.01 & 1.65 &  1.63& 1.98 & 2.18 & 1.61 & 2.16 & 1.60\\ \hline
\multicolumn{9}{l}{$\dagger$ All other source parameters have been kept unchanged.}

\end{tabular}
\label{testonmodels}
\end{center}
\end{table*}



\bsp	
\label{lastpage}
\end{document}